\documentstyle[12pt,epsf,citesort]{article}
\newcommand{\Z}{{\sf Z \!\!\! Z}}
\newcommand{\R}{{\sf I \!\! R}}
\newcommand{\1}{{\sf 1 \!\! 1}}

\newcommand{\p}{\partial}
\newcommand{\e}{\vec e}

\newcommand{\pAP}{\psi_+^A}
\newcommand{\pAPD}{\psi_+^{A\dagger}}
\newcommand{\pBP}{\psi_+^B}
\newcommand{\pBPD}{\psi_+^{B\dagger}}
\newcommand{\pAM}{\psi_-^A}
\newcommand{\pAMD}{\psi_-^{A\dagger}}
\newcommand{\pBM}{\psi_-^B}
\newcommand{\pBMD}{\psi_-^{B\dagger}}
\makeatletter
\@addtoreset{equation}{section}
\makeatother

\title{Systematic Low-Energy Effective Theory for Magnons and Charge Carriers
in an Antiferromagnet}

\author{F.~K\"ampfer, M.~Moser, and U.-J.~Wiese
\\ \\
Institute for Theoretical Physics, Bern University \\
Sidlerstrasse 5, CH-3012 Bern, Switzerland}

\begin{document} 
\maketitle

\vspace{-1cm}

\begin{abstract} \normalsize

By electron or hole doping quantum antiferromagnets may turn into 
high-temperature superconductors. The low-energy dynamics of antiferromagnets 
are governed by their Nambu-Goldstone bosons --- the magnons --- and are 
described by an effective field theory analogous to chiral perturbation theory 
for the pions in strong interaction physics. In analogy to baryon chiral 
perturbation theory --- the effective theory for pions and nucleons ---  we 
construct a systematic low-energy effective theory for magnons and electrons or
holes in an antiferromagnet. The effective theory is universal and makes 
model-independent predictions for the entire class of antiferromagnetic 
cuprates. We present a detailed analysis of the symmetries of the Hubbard model
and discuss how these symmetries manifest themselves in the effective theory. 
A complete set of linearly independent leading contributions to the effective
action is constructed. The coupling to external electromagnetic fields is also 
investigated. 

\end{abstract}
 
\maketitle
 
\newpage

\section{Introduction}

Almost 20 years after the discovery of high-temperature superconductivity in 
layered cuprates \cite{Bed86}, identifying the dynamical mechanism behind it
remains one of the great challenges in condensed matter physics. Ordinary 
low-temperature superconductors are weakly coupled electron systems in which 
phonon exchange mediates an attractive interaction that can overcome the 
Coulomb repulsion between electrons. As massless Nambu-Goldstone bosons of the 
spontaneously broken translation symmetry, phonons provide a natural mechanism 
for Cooper pair formation at low energies which is successfully quantified in 
BCS theory. In contrast to ordinary superconductors, layered high-$T_c$ 
cuprates are systems of strongly correlated electrons to which the weak 
coupling BCS theory is not readily applicable. Furthermore, the high transition
temperatures of cuprate superconductors and the smallness of the isotope effect
suggest that mechanisms other than phonon exchange may be responsible for 
Cooper pair formation. Since high-temperature superconductors are 
antiferromagnets before doping, it is natural to suspect (but not generally 
accepted) that magnons --- the Nambu-Goldstone bosons of the spontaneously 
broken $SU(2)_s$ spin symmetry --- may be important for binding electrons or 
holes into preformed pairs. 

Even if spin fluctuations were not the key to explaining high-temperature 
superconductivity, the dynamics of charge carriers in an antiferromagnet is an 
interesting topic in itself. There is a vast literature on this subject. The 
dynamics of holes in an antiferromagnet has been investigated, for example, in
\cite{Bri70,Hir85,And87,Gro87,Shr88,Tru88,Sch88,Kan89,Sac89,Wen89,Kra89,Sha90,And90,Sin90,Tru90,Els90,Dag90,Vig90,Kop90,Ver91,Mar91,Mon91,Liu92,Kue93,Dah93,Sus94,Dag94,Alt95,Kyu97,Chu98,Kar98,Ape98,Bru00,Sac03,Liu03}.
Understanding the dynamics of even just a single hole propagating in an 
antiferromagnet is a challenging problem. One can gain qualitative insight from
a picture in which holes hop from site to site, leaving a string 
of flipped spins behind and thus locally destroying the antiferromagnetic 
order. Since the string costs energy proportional to its length, one might 
expect the holes to even be confined and thus have infinite mass. However, the 
locally destroyed antiferromagnetic order may be healed by appropriate hole 
hopping which renders the hole mass finite \cite{Tru88}. Angle resolved 
photoemission spectroscopy experiments \cite{Wel95,LaR97,Kim98,Ron98} as well 
as a number of theoretical investigations \cite{Tru88,Shr88,Els90,Bru00}
indicate that the minimum of the dispersion (i.e.\ of the energy) of a single 
hole corresponds to lattice momenta $(\pm \frac{\pi}{2},\pm \frac{\pi}{2})$ in 
the Brillouin zone.

As one adds a second hole, the situation becomes more controversial. For 
example, there seems to be no consensus on the question if a pair of holes can
form a bound state or not. If it can, the condensation of such pairs would 
provide a potential mechanism for high-temperature superconductivity. The 
effective theory to be constructed here can be used to analytically calculate 
the long-range magnon-mediated forces between holes using perturbation theory. 
It is very interesting to ask what happens when one dopes an antiferromagnet 
with a non-zero density of holes. At sufficient doping, experiments show that 
high-temperature superconductivity may arise. It has been argued on theoretical
grounds that even an infinitesimal amount of doping may affect the 
antiferromagnetic phase and turn it into a spiral phase \cite{Shr88}. A
systematic investigation of this question is also possible using the effective 
theory of this paper, but it will require the use of non-perturbative methods.

The standard models for antiferromagnets and high-temperature superconductors 
are the Hubbard and $t$-$J$ model. Since these models are strongly coupled,
they are not accessible to a systematic analytic treatment. As a consequence, 
analytic calculations in Hubbard-type models usually involve some uncontrolled 
approximations. Unfortunately, due to a severe fermion sign problem, away from 
half-filling these models can currently also not be simulated reliably. Hence, 
although they may indeed contain the relevant physics, Hubbard-type models have
not yet led to a quantitative understanding of high-$T_c$ materials. An 
alternative to a microscopic description using Hubbard-type models is provided 
by phenomenological models formulated directly in terms of magnon and electron 
or hole fields \cite{Wen89,Sha90,Kue93,Kar98,Liu03}. Although they may provide 
qualitative insight, such models do not lead to unambiguous predictions. In 
this paper, for the first time we introduce a systematic low-energy effective 
field theory for magnons and charge carriers in an antiferromagnet. Based only 
on symmetries and their spontaneous breakdown, the effective theory makes 
universal predictions for the entire class of antiferromagnetic cuprates. 
Although the effective theory is not renormalizable, it yields unambiguous 
results in a systematic low-energy expansion. In each order of the expansion, 
the results depend only on a finite number of material specific low-energy 
parameters whose values can be determined experimentally. The effective theory 
is not based on a specific microscopic model Hamiltonian but is universally 
applicable. Furthermore, and most important, in contrast to the strongly 
correlated electrons of Hubbard-type models, the electrons and holes of the 
effective field theory are quasi-particles that are weakly coupled to the 
magnons. Consequently, one may expect that the effective theory is more easily 
solvable than the underlying microscopic models.

Possible basic applications of the effective theory to be constructed in this 
paper include magnon-magnon, magnon-hole, and magnon-electron scattering as 
well as the determination of long-range magnon-mediated forces between the 
charge carriers. More ambitious applications could aim at a quantitative 
explanation of the Mott insulator state, the reduction of the staggered 
magnetization upon doping, the formation of a spiral phase, or at a systematic 
study of potential mechanisms for the preformation of electron or hole pairs in
the antiferromagnetic phase. When such pairs condense they may become the 
Cooper pairs of high-temperature superconductivity. Except for a derivation of 
the dispersion relation of charge carriers, in this paper we do not consider 
applications yet, but concentrate entirely on the construction of the effective
theory itself.

The construction in this paper is inspired by similar developments in the
theory of the strong interactions. In contrast to the high-$T_c$ problem, where
the choice of a microscopic model is controversial, there is general agreement 
that Quantum Chromodynamics (QCD) provides the correct microscopic description 
of the strong interactions. Still, similar to Hubbard-type models, solving
QCD is notoriously hard. At ``half-filling'', i.e.\ in the filled quark Dirac 
sea that represents 
the QCD vacuum, the $SU(2)_L \otimes SU(2)_R$ chiral symmetry of massless up 
and down quarks is spontaneously broken to the isospin symmetry $SU(2)_{L=R}$, 
resulting in three massless Nambu-Goldstone pions. This is analogous to the 
spontaneous breaking of the $SU(2)_s$ spin symmetry down to $U(1)_s$ that leads
to antiferromagnetism. The corresponding Nambu-Goldstone bosons --- in this 
case two magnons  --- are thus analogous to the pions of the strong 
interactions. It is possible to study chiral symmetry breaking in the QCD 
vacuum in numerical simulations of lattice QCD, just as it is possible to study
antiferromagnetism by simulating the Hubbard model at half-filling. However, it
is very useful to also investigate these phenomena with effective field 
theories. The low-energy effective theory for pions was pioneered by Weinberg
\cite{Wei79} and formulated as a systematic expansion in Gasser's and 
Leutwyler's chiral perturbation theory \cite{Gas85}. Based on symmetry 
considerations and the observation that chiral symmetry is spontaneously 
broken, chiral perturbation theory makes rigorous predictions about the pion 
dynamics in terms of a few low-energy parameters such as the pion decay 
constant, the chiral condensate, and the Gasser-Leutwyler coefficients. Once 
these parameters are determined, either experimentally or through lattice QCD 
calculations, the effective theory makes unambiguous predictions in the 
low-energy domain. 

Chiral perturbation theory can be applied to any Nambu-Goldstone 
phe\-no\-me\-non, and has indeed been used for both ferro- \cite{Leu94,Hof99} 
and antiferromagnetic magnons \cite{Neu89,Fis89,Has90,Has93,Chu94,Rom99}. To 
lowest order, for antiferromagnetic magnons the low-energy parameters of chiral
perturbation theory are the spin stiffness $\rho_s$ and the spin-wave velocity 
$c$. At low energies chiral perturbation theory describes all aspects of the 
magnon 
dynamics just in terms of these two parameters. For example, the low-energy 
physics of the Hubbard model at half-filling is completely described by the 
effective theory once $\rho_s$ and $c$ have been determined in terms of the 
Hubbard model parameters $t$ and $U$.

A numerical challenge in high-$T_c$ physics is to simulate the Hubbard model 
away from half-filling. This requires a solution of the corresponding fermion 
sign problem. Similarly, simulating lattice QCD at non-zero baryon chemical 
potential, i.e.\ after ``doping'' the QCD vacuum with quarks, is prevented by a
severe complex action problem. Like for high-$T_c$ materials at sufficient
doping, one expects that QCD at sufficiently high baryon density becomes a 
superconductor, in that case for the color charge carried by quarks and gluons 
\cite{Raj00}. In contrast to high-temperature superconductivity, the mechanism 
responsible for color-superconductivity is well understood in terms of 
one-gluon exchange. Color-superconductivity requires very large baryon 
densities and may thus arise only in the core of compact neutron or quark 
stars. However, superconductivity --- not of color but of ordinary electric 
charge --- is also known to exist at more moderate baryon densities. In 
particular, pairing of protons or neutrons inside large nuclei or neutron stars
leads to superconductivity or superfluidity. Understanding the mechanism of 
nucleon pairing from the microscopic QCD theory may be as hard as understanding
the mechanism for high-temperature superconductivity directly from the Hubbard 
model. Instead it is much more useful to employ a systematic low-energy 
effective theory whose parameters can be determined from the underlying 
microscopic physics. In nuclear physics effective field theory has recently led
to some progress in describing the forces between nucleons in terms of just a 
few low-energy parameters 
\cite{Wei90,Kap98,Epe98,Bed98,Kol99,Par99,Epe01,Bea02,Bed02,Nog05}, 
while phenomenological models involve a much larger number of adjustable 
parameters. Also steps towards describing nuclear matter with effective field
theories have already been taken \cite{Rho00,Par00,Mue00,Mei02,Wir03}. The goal
of the present paper is to develop a similar effective theory describing the 
interactions between the charge carriers in an antiferromagnet through magnon 
exchange. Remarkably, some physical phenomena that are practically inaccessible
to microscopic Hubbard-type models even by numerical simulation can be tackled 
analytically in the effective field theory framework.

An ambitious goal of the effective theory approach is to systematically
investigate possible mechanisms for the preformation of electron or hole pairs 
as a potential step towards understanding high-temperature superconductivity. 
It is an experimental fact that antiferromagnetism is destroyed before one 
enters the superconducting phase. How can magnon exchange then possibly provide
a mechanism relevant for Cooper pair preformation? The destruction of 
antiferromagnetism just means the absence of infinite-range antiferromagnetic 
order. Antiferromagnetic correlations, although only of finite range, exist 
even in the superconducting phase. The finite correlation length implies that 
the magnons have developed a massgap, but they may still exist as relevant 
low-energy degrees of freedom. In particular, in $2+1$ dimensions, as a 
consequence of the Hohenberg-Mermin-Wagner-Coleman theorem, magnons pick up a 
mass that is exponentially small in the inverse temperature \cite{Cha89,Has91}.
The generation of the massgap is a non-perturbative phenomenon that is well 
within the applicability range of the effective theory, although infinite-range
antiferromagnetic order exists only at zero temperature. Similarly, an 
effective theory for magnons and electrons or holes remains valid in the 
superconducting phase as long as the magnons remain among the lightest degrees 
of freedom. Again, this is similar to QCD where pions are not exactly massless 
either --- in that case as a result of explicit chiral symmetry breaking due to
non-zero quark masses. Although pions are hence only pseudo-Nambu-Goldstone 
bosons, chiral perturbation theory remains perfectly well applicable.

The low-energy effective theory for magnons and charge carriers to be 
developed here is the condensed matter analog of baryon chiral perturbation 
theory in strong interaction physics \cite{Geo84a,Gas88,Jen91,Ber92,Bec99}. The
effective theory is based on a non-linear realization of the spontaneously 
broken symmetry \cite{Col69,Cal69}. The terms in the low-energy effective 
Lagrangian are organized according to the number of derivatives they contain. 
The lowest energy physics is dominated by the terms with the smallest number of
derivatives, while effects at higher energies are taken into account 
systematically through higher-derivative terms. A key ingredient in 
constructing the effective Lagrangian are symmetry considerations. At a given 
order of the low-energy expansion, i.e.\ for a given number of derivatives, all
terms consistent with the symmetries must be included in the effective 
Lagrangian, with a low-energy parameter that determines the strength of the 
corresponding interaction. For cuprates the most important symmetries are the 
$SU(2)_s$ spin symmetry which is spontaneously broken down to $U(1)_s$ in the 
antiferromagnetic phase, as well as the $U(1)_Q$ fermion number symmetry whose 
breakdown signals superconductivity. Other relevant symmetries include 
translation by one lattice spacing which changes the sign of the staggered 
magnetization, 90 degrees rotations and reflections of the square crystal 
lattice, as well as time-reversal. In addition to these generic symmetries of 
high-$T_c$ materials, the Hubbard model possesses an $SU(2)_Q$ symmetry 
discussed by Yang and Zhang \cite{Zha90,Yan90} which is a non-Abelian 
extension of the charge symmetry $U(1)_Q$. This symmetry is not expected to be 
present in generic cuprate materials, but may still be a relevant approximate
symmetry in specific samples. 

In this paper we ignore phonons, assuming that they do not play an important 
role for high-temperature superconductivity. For example, in the Hubbard model 
a rigid lattice which does not have its own physical degrees of freedom is put 
by hand. Of course, in the actual high-$T_c$ materials a crystal lattice arises
as a result of the spontaneous breakdown of translation and Galilean (or more
precisely Poincar\'e) invariance. The corresponding Nambu-Goldstone bosons
are the phonons. The role of phonons and their possible interplay with magnons 
can also be investigated systematically in the framework of low-energy 
effective field theory.

We also consider the coupling of antiferromagnets to external 
electromagnetic fields which can be used to probe the dynamics of magnons and
electrons or holes. As first noted by Fr\"ohlich and Studer, in 
non-relativistic condensed matter external electromagnetic fields $\vec E(x)$ 
and $\vec B(x)$ enter the dynamics in the form of non-Abelian vector potentials
for the $SU(2)_s$ spin symmetry \cite{Fro92}. We use this observation to couple
both the microscopic Hubbard model and the effective theory to external 
$\vec E(x)$ and $\vec B(x)$ fields. As discussed in detail in \cite{Bae04}, the
electromagnetic couplings are the condensed matter analog of the weak 
interactions in particle physics. These are described by an 
$SU(2)_L \otimes U(1)_Y$ gauge theory, which turns part of QCD's global chiral 
symmetry into a gauge symmetry. Remarkably, the electromagnetic couplings of 
non-relativistic condensed matter are described by a local 
$SU(2)_s \otimes U(1)_Q$ symmetry which is the condensed matter analog of the 
$SU(2)_L \otimes U(1)_Y$ symmetry in particle physics. Some correspondences 
between QCD and antiferromagnets are summarized in table 1. Connections between
QCD and condensed matter physics have also been discussed in \cite{Cha03a}.

\begin{table}[t]
\begin{center}
\begin{tabular}{|c|c|c|}
\hline
 & QCD & Antiferromagnets \\
\hline
\hline
broken phase & hadronic vacuum & antiferromagnetic phase \\
\hline
global symmetry & chiral symmetry & spin rotations \\
\hline
symmetry group $G$ & $SU(2)_L \otimes SU(2)_R$ & $SU(2)_s$ \\
\hline
unbroken subgroup $H$ & $SU(2)_{L=R}$ & $U(1)_s$ \\
\hline
Goldstone boson & pion & magnon \\
\hline
Goldstone field in $G/H$ & $U(x) \in SU(2)$ & $\e(x) \in S^2$ \\
\hline
order parameter & chiral condensate & staggered magnetization \\
\hline
coupling strength & pion decay constant $F_\pi$ & spin stiffness $\rho_s$ \\
\hline
propagation speed & velocity of light & spin-wave velocity \\
\hline
conserved charge & baryon number $U(1)_B$ & electric charge $U(1)_Q$ \\
\hline
charged particle & nucleon or antinucleon & electron or hole \\
\hline
long-range force & pion exchange & magnon exchange \\
\hline
weak probes & electroweak fields & electromagnetic fields \\
\hline
local symmetry of & electroweak & local \\
weak probes & $SU(2)_L \otimes U(1)_Y$ & $SU(2)_s \otimes U(1)_Q$ \\
\hline
dense phase & nuclear or quark matter & high-$T_c$ superconductor \\
\hline
microscopic description & lattice QCD & Hubbard-type models \\
\hline
effective description & chiral perturbation & magnon effective \\
of Goldstone bosons & theory & theory \\
\hline
effective description & baryon chiral & effective theory \\
of charged fields & perturbation theory & presented here \\
\hline
\end{tabular}
\end{center}
\caption{\it Correspondences between QCD and antiferromagnets.}
\end{table}

The rest of this paper is organized as follows. Section 2 contains a symmetry
analysis of the Hubbard model as a concrete example for an underlying
microscopic system. In section 3 the effective theory for magnons is reviewed
and the non-linear realization of the $SU(2)_s$ spin symmetry is constructed.
In section 4 the Hubbard model is coupled to a magnon background field. In this
way the fields of the effective theory inherit their transformation properties 
under the various symmetries from the underlying microscopic degrees of 
freedom. In section 5 the low-energy effective theory for magnons and charge 
carriers is developed and the leading terms in a systematic low-energy 
expansion of the effective action are constructed. This section also contains 
an application of the effective theory to the dispersion relations of charge 
carriers. Section 6 treats the $t$-$J$ model and its effective theory as a
special case of systems with holes as the only charge carriers. In section 7 
the Hubbard model as well as its effective theory are coupled to external 
electromagnetic fields. Finally, section 8 contains our conclusions, while some
technical details are discussed in two appendices.

\section{Symmetries of the Hubbard Model}

In order to have a concrete microscopic system for which we will then
construct a low-energy effective theory, we consider the Hubbard model. The
Hubbard model just serves as one representative of a large class of systems,
including the actual high-$T_c$ materials. Here it is essential that the 
Hubbard model shares important symmetries, e.g.\ an $SU(2)_s$ spin symmetry
and a $U(1)_Q$ fermion number symmetry with these materials. In the Hubbard 
model at half-filling the $U(1)_Q$ symmetry even extends to an $SU(2)_Q$ 
symmetry. The $SU(2)_Q$ symmetry is not exact in actual materials, but may 
still be approximately realized and will also be investigated in the framework
of the effective theory.

\subsection{Hamiltonian and Generic Continuous Symmetries}

The Hubbard model is defined by the Hamiltonian
\begin{eqnarray}
H&=&- t \sum_{x, i} 
(c_{x \uparrow}^\dagger c_{x+\hat i \uparrow} + 
c_{x + \hat i \uparrow}^\dagger c_{x \uparrow} +
c_{x \downarrow}^\dagger c_{x+\hat i \downarrow} + 
c_{x + \hat i \downarrow}^\dagger c_{x \downarrow}) \nonumber \\
&&+ U \sum_x c_{x \uparrow}^\dagger c_{x \uparrow}
c_{x \downarrow}^\dagger c_{x \downarrow} - 
\mu' \sum_x (c_{x \uparrow}^\dagger c_{x \uparrow} +
c_{x \downarrow}^\dagger c_{x \downarrow}).
\end{eqnarray}
Here $x$ denotes the sites of a 2-dimensional square lattice and $\hat i$ is a 
vector of length $a$ (where $a$ is the lattice spacing) pointing in the 
$i$-direction. Furthermore, $t$ is the nearest-neighbor hopping parameter, 
while $U>0$ is the strength of the screened on-site Coulomb repulsion, and 
$\mu'$ is the chemical potential for fermion number. The fermion creation and 
annihilation operators obey the standard anticommutation relations
\begin{equation}
\{c_{x s}^\dagger,c_{y s'}\} = \delta_{xy} \delta_{s s'}, \quad
\{c_{x s},c_{y s'}\} = \{c_{x s}^\dagger,c_{y s'}^\dagger\} = 0.
\end{equation}
We also introduce the $SU(2)_s$ Pauli spinor
\begin{equation}
c_x = \left(\begin{array}{c} c_{x \uparrow} \\ c_{x \downarrow} 
\end{array} \right)
\end{equation}
in terms of which (up to an irrelevant constant) the Hamiltonian takes the 
manifestly $SU(2)_s$-invariant form
\begin{equation}
H = - t \sum_{x, i} (c_x^\dagger c_{x+\hat i} + 
c_{x + \hat i}^\dagger c_x) + 
\frac{U}{2} \sum_x (c_x^\dagger c_x - 1)^2 - \mu \sum_x (c_x^\dagger c_x - 1).
\end{equation}
Here $\mu = \mu' - \frac{1}{2} U$ is the chemical potential for the fermion 
number relative to half-filling, i.e.\ $\mu = 0$ implies an average density of 
one fermion per lattice site. The corresponding $U(1)_Q$ symmetry is generated 
by the charge operator
\begin{equation}
Q = \sum_x Q_x = \sum_x (c_x^\dagger c_x - 1).
\end{equation}
Again, we count fermion number relative to half-filling. The $SU(2)_s$ symmetry
is generated by the total spin
\begin{equation}
\label{spin}
\vec S = \sum_x \vec S_x = 
\sum_x c_x^\dagger \ \frac{\vec \sigma}{2} \ c_x,
\end{equation}
where $\vec \sigma$ are the Pauli matrices. It is easy to see that the above 
Hamiltonian conserves both fermion number and spin, i.e.\ 
$[H,Q] = [H,\vec S] = 0$, and that $[Q,\vec S] = 0$. The infinitesimal 
generators $\vec S$ of $SU(2)_s$ (which obey the standard commutation relations
$[S_a,S_b] = i \varepsilon_{abc} S_c$) can be used to construct a unitary 
operator
\begin{equation}
V = \exp(i \vec \eta \cdot \vec S),
\end{equation}
which implements the corresponding symmetry transformations in the Hilbert
space of the theory. In particular, the transformed annihilation
operators take the form
\begin{equation}
\label{trafog}
c_x' = V^\dagger c_x V = \exp(i \vec \eta \cdot \frac{\vec \sigma}{2}) c_x = 
g c_x, \quad g = \exp(i \vec \eta \cdot \frac{\vec \sigma}{2}) \in SU(2)_s.
\end{equation}
Similarly, the $U(1)_Q$ transformations are implemented by a unitary operator
\begin{equation}
W = \exp(i \omega Q),
\end{equation}
such that
\begin{equation}
^Qc_x = W^\dagger c_x W = \exp(i \omega) c_x, \quad \exp(i \omega) \in U(1)_Q.
\end{equation}

For large positive $U$, at half-filling, the repulsive Hubbard model reduces to
the antiferromagnetic spin $\frac{1}{2}$ quantum Heisenberg model with the 
Hamiltonian
\begin{equation}
\label{Heisenberg}
H = J \sum_{x,i} \vec S_x \cdot \vec S_{x+\hat i},
\end{equation}
where the exchange coupling is given by $J = 2 t^2/U$. This follows to second 
order of perturbation theory in $t/U$. To leading order, i.e.\ completely 
ignoring the kinetic term proportional to $t$, there is an enormous number of
degenerate ground states. Irrespective of spin, any state with exactly one 
fermion occupying each lattice site avoids the on-site Coulomb repulsion and 
thus represents a ground state for $t=0$. There is no correction at order 
$t/U$. In second order of degenerate perturbation theory, a spin can virtually 
hop to a neighboring site occupied by a fermion with opposite spin and then hop
back. On the other hand, virtual hops to sites occupied by a fermion with the 
same spin orientation are forbidden by the Pauli principle. This favors 
antiparallel spins and leads to the antiferromagnetic Heisenberg model of 
eq.(\ref{Heisenberg}).

\subsection{Discrete Symmetries}

Since the Hubbard model at half-filling leads to antiferromagnetism, another 
important symmetry is translation by one lattice spacing (in the 
$i$-direction), which flips the sign of the staggered magnetization vector
\begin{equation}
\vec M_s = \sum_x (-1)^x \vec S_x.
\end{equation}
The factor $(-1)^x = (-1)^{(x_1+x_2)/a}$ distinguishes between the sites of the
even and odd sublattice. The points on the even sublattice $A$ have 
$(-1)^x = 1$ while the points on the odd sublattice $B$ have $(-1)^x = - 1$. 
The displacement symmetry is generated by a unitary operator $D$ which acts as
\begin{equation}
^Dc_x = D^\dagger c_x D = c_{x+\hat i},
\end{equation}
and for which $[H,D] = 0$. Obviously, both the $U(1)_Q$ and the $SU(2)_s$ 
symmetry commute with the displacement, i.e.\ $[Q,D] = [\vec S,D] = 0$. In the 
effective theory it will be useful to also consider a related symmetry $D'$ 
which combines $D$ with the spin rotation $g = i \sigma_2$. This symmetry 
acts as
\begin{equation}
^{D'}c_x = D'^\dagger c_x D' = (i \sigma_2)\ ^Dc_x = (i \sigma_2) c_{x+\hat i}.
\end{equation}
Also note that $[H,D'] = [D,D'] = [Q,D'] = 0$, but $[\vec S,D'] \neq 0$.

In non-relativistic physics orbital angular momentum and spin are separately
conserved and spin plays the role of an internal quantum number. Indeed, in the
Hubbard model the $SU(2)_s$ spin symmetry is completely independent of the
90 degrees rotation invariance of the spatial lattice. The 90 degrees rotation 
$O$ acts on a spatial point $x = (x_1,x_2)$ as $Ox = (- x_2,x_1)$. Under the 
symmetry $O$ the fermion operators transform as 
\begin{equation}
^Oc_x = O^\dagger c_x O = c_{Ox}. 
\end{equation}
Parity turns $x$ into $(- x_1,- x_2)$ and is equivalent to a 180 degrees 
rotation in two dimensions. Hence, it is more useful to consider the spatial 
reflection $R$ at the $x_1$-axis which turns $x$ into $Rx = (x_1,- x_2)$. Under
this transformation the fermion operators transform as 
\begin{equation}
^Rc_x = R^\dagger c_x R = c_{Rx}.
\end{equation} 
The reflection at the orthogonal $x_2$-axis is a combination of the reflection 
$R$ and the rotation $O$. One can also consider the reflection at an axis half 
between lattice points. This transformation is a combination of $R$ with the 
displacement symmetry $D$. Similarly, a reflection at a lattice diagonal is a 
combination of $R$ and $O$. Another important symmetry is time-reversal which 
is implemented by an antiunitary operator $T$.

It should be pointed out that, unlike the actual high-$T_c$ materials, the 
Hubbard model is not Galilean invariant: in the actual materials translation as
well as Galilean invariance are spontaneously broken by the formation of the 
crystal lattice. The corresponding Nambu-Goldstone bosons are the phonons which
are known to play a central role in ordinary low-$T_c$ superconductivity. In
the Hubbard model, on the other hand, the lattice is imposed by hand, and thus 
translation and Galilean invariance are explicitly broken. In particular, 
phonons cannot arise because the lattice does not have its own physical degrees
of freedom.

\subsection{$SU(2)_Q$ Symmetry}

As first noted by Yang and Zhang \cite{Zha90,Yan90}, at half-filling (i.e.\ for
$\mu = 0$) the Hubbard model possesses a non-Abelian extension $SU(2)_Q$ of the
fermion number symmetry $U(1)_Q$ generated by 
\begin{equation}
Q^+ = \sum_x (-1)^x c_{x \uparrow}^\dagger c_{x \downarrow}^\dagger, \
Q^- = \sum_x (-1)^x c_{x \downarrow} c_{x \uparrow}, \
Q^3 = \sum_x \frac{1}{2}(c_{x \uparrow}^\dagger c_{x \uparrow} +
c_{x \downarrow}^\dagger c_{x \downarrow} - 1) = \frac{1}{2} Q.
\end{equation}
Writing $Q^\pm = Q^1 \pm i Q^2$, it is straightforward to show that, for 
$\mu = 0$, indeed $[H,\vec Q] = 0$. Also the $SU(2)_Q$ symmetry commutes with 
the $SU(2)_s$ symmetry, i.e.\ $[Q^a,S^b] = 0$, but it does not commute with the
displacement symmetry because $D^\dagger Q^\pm D = - Q^\pm$. For the same 
reason $[\vec Q,D'] \neq 0$.

Introducing the $SU(2)_Q$ spinor
\begin{equation}
d_x = \left(\begin{array}{c} c_{x \uparrow} \\ 
(-1)^x c_{x \downarrow}^\dagger \end{array} \right),
\end{equation}
which obeys the standard anticommutation relations
\begin{equation}
\{d_{x a}^\dagger,d_{y b}\} = \delta_{xy} \delta_{ab}, \quad
\{d_{x a},d_{y b}\} = \{d_{x a}^\dagger,d_{y b}^\dagger\} = 0,
\end{equation}
one writes
\begin{equation}
\vec Q = \sum_x \vec Q_x = 
\sum_x d_x^\dagger \ \frac{\vec \sigma}{2} \ d_x,
\end{equation}
where $\vec \sigma$ are again the Pauli matrices now operating in $SU(2)_Q$
space. The infinitesimal generators $\vec Q$ of $SU(2)_Q$ can be used to 
construct a unitary operator
\begin{equation}
W = \exp(i \vec \omega \cdot \vec Q),
\end{equation}
which implements the corresponding symmetry transformations in the Hilbert
space of the theory. The transformed $SU(2)_Q$ spinors are then given by
\begin{equation}
^{\vec Q}d_x = W^\dagger d_x W = 
\exp(i \vec \omega \cdot \frac{\vec \sigma}{2}) d_x = \Omega d_x, \quad
\Omega = \exp(i \vec \omega \cdot \frac{\vec \sigma}{2}) \in SU(2)_Q.
\end{equation}
In terms of the $SU(2)_Q$ spinors the Hamiltonian takes the form
\begin{equation}
\label{Hd}
H = - t \sum_{x, i} (d_x^\dagger d_{x+\hat i} + 
d_{x + \hat i}^\dagger d_x) - \frac{U}{2} \sum_x (d_x^\dagger d_x - 1)^2 -
\mu \sum_x d_x^\dagger \sigma_3 d_x.
\end{equation}
The first two terms on the right-hand side are manifestly $SU(2)_Q$-invariant,
while away from half-filling (i.e.\ for $\mu \neq 0$) the chemical potential 
term explicitly breaks the $SU(2)_Q$ symmetry down to $U(1)_Q$.

Finally, we introduce a matrix-valued fermion operator
\begin{equation}
\label{Coperator}
C_x = \left(\begin{array}{cc} c_{x \uparrow} &
(-1)^x c^\dagger_{x \downarrow} \\ c_{x \downarrow} &
- (-1)^x c_{x \uparrow}^\dagger \end{array} \right),
\end{equation}
which displays both the $SU(2)_s$ and the $SU(2)_Q$ symmetries in a compact 
form. The first column of $C_x$ is the $SU(2)_s$ spinor $c_x$, while the 
second column is another $SU(2)_s$ spinor which transforms exactly like $c_x$. 
The first row of $C_x$ is the $SU(2)_Q$ spinor $d_x^T$, while the second row is
another $SU(2)_Q$ spinor which transforms exactly like $d_x^T$. Under combined 
$SU(2)_s$ and $SU(2)_Q$ transformations $C_x$ transforms as
\begin{equation}
^{\vec Q}C_x' = g C_x \Omega^T.
\end{equation}
Since the $SU(2)_s$ symmetry acts on the left while the $SU(2)_Q$ symmetry acts
on the right, it is now manifest that the two symmetry operations commute.
Under the displacement symmetry one obtains
\begin{equation}
^DC_x = C_{x+\hat i} \sigma_3.
\end{equation}
The appearance of $\sigma_3$ on the right is due to the factor $(-1)^x$ and
confirms that the displacement symmetry commutes with all $SU(2)_s$ 
transformations, but only with the Abelian $U(1)_Q$ (and not with all 
$SU(2)_Q$) transformations. Similarly, under the symmetry $D'$ one finds
\begin{equation}
^{D'}C_x = (i \sigma_2) C_{x+\hat i} \sigma_3.
\end{equation}
The Hamiltonian can now be expressed in a manifestly $SU(2)_s$-, $U(1)_Q$-, 
$D$-, and $D'$-invariant form
\begin{equation}
\label{HF}
H = - \frac{t}{2} \sum_{x, i} \mbox{Tr}[C_x^\dagger C_{x+\hat i} +
C_{x+\hat i}^\dagger C_x] + 
\frac{U}{12} \sum_x \mbox{Tr}[C_x^\dagger C_x C_x^\dagger C_x] -
\frac{\mu}{2} \sum_x \mbox{Tr}[C_x^\dagger C_x \sigma_3].
\end{equation}
The chemical potential term is only $U(1)_Q$ invariant, while the other two 
terms are manifestly $SU(2)_Q$-invariant.

\section{Effective Theory for Magnons}

Before doping, the high-$T_c$ materials are quantum antiferromagnets in which 
the $SU(2)_s$ spin symmetry is spontaneously broken down to $U(1)_s$. The
low-energy physics of antiferromagnets is dominated by the corresponding
Nambu-Gold\-stone bosons --- the magnons. Chiral perturbation theory, which was
originally developed for the Nambu-Goldstone pions of QCD, is a systematic
low-energy expansion that has also been applied to magnons 
\cite{Neu89,Fis89,Has90,Has93,Leu94,Hof99,Chu94,Rom99}. In this section
we review the basic features of magnon chiral perturbation theory. As a
necessary prerequisite for the coupling of magnons to charge carriers, we 
also construct the non-linear realization of a spontaneously broken $SU(2)_s$
symmetry, which then appears as a local $U(1)_s$ symmetry in the unbroken 
subgroup. This is analogous to baryon chiral perturbation theory in which the 
spontaneously broken $SU(2)_L \otimes SU(2)_R$ chiral symmetry of QCD is 
implemented on the nucleon fields as a local $SU(2)_{L=R}$ transformation in 
the unbroken isospin subgroup.

\subsection{Continuous Symmetries of the Effective Action}

The undoped precursors of high-temperature layered cuprate superconductors are 
quantum antiferromagnets. At half-filling, also the Hubbard model displays 
antiferromagnetism. In these systems, at least at zero temperature, the spin 
rotational symmetry $G = SU(2)_s$ is spontaneously broken down to the subgroup 
$H = U(1)_s$ by the formation of a staggered magnetization. The $U(1)_Q$
symmetry, on the other hand, remains unbroken until one reaches the 
superconducting phase. In the Hubbard model even the $SU(2)_Q$ symmetry remains
unbroken at half-filling but is explicitly broken down to $U(1)_Q$ for
$\mu \neq 0$. As a consequence of Goldstone's theorem, there are two massless 
bosons --- the antiferromagnetic spin-waves or magnons, which are described by 
a unit-vector field 
\begin{equation}
\e(x) = (e_1(x),e_2(x),e_3(x)), \quad \e(x)^2 = 1
\end{equation}
in the coset space $G/H = SU(2)_s/U(1)_s = S^2$. Here $x = (x_1,x_2,t)$ denotes
a point in Euclidean space-time. The vector $\vec e(x)$ describes the direction
of the local staggered magnetization. The leading order terms in the Euclidean 
action of the low-energy effective theory for the magnons take the form 
\cite{Cha89,Has91}
\begin{equation}
\label{action}
S[\e] = \int d^2x \ dt \ \frac{\rho_s}{2} 
\left(\p_i \e \cdot \p_i \e + \frac{1}{c^2} \p_t \e \cdot \p_t \e\right).
\end{equation}
The index $i \in \{1,2\}$ labels the two spatial directions, while the index 
$t$ refers to the Euclidean time-direction. The parameter $\rho_s$ is the spin 
stiffness and $c$ is the spin-wave velocity. For the antiferromagnetic 
Heisenberg model of eq.(\ref{Heisenberg}) these low-energy parameters have been
determined very precisely in Monte Carlo calculations \cite{Wie94,Bea96} 
resulting in $\rho_s = 0.186(4) J$, $c = 1.68(1) J a$, where $J$ is the 
exchange coupling of the Heisenberg model and $a$ is the lattice spacing. The 
leading terms in the magnon effective action are ``Poincar\'e''-invariant with 
the spin-wave velocity $c$ playing the role of the velocity of light. 
Consequently, antiferromagnetic magnons have a ``relativistic'' spectrum. The 
``Poincar\'e'' symmetry emerges only at low energies as a consequence of the 
discrete lattice rotation invariance. However, higher-derivative terms relevant
at higher energies are in general not invariant.

In the following we prefer to work with an alternative representation of the
magnon field using $2 \times 2$ Hermitean projection matrices $P(x)$ that obey
\begin{equation}
P(x)^\dagger = P(x), \quad \mbox{Tr} P(x) = 1, \quad P(x)^2 = P(x),
\end{equation}
and are given by
\begin{equation}
\label{CP1O3}
P(x) = \frac{1}{2}(\1 + \e(x) \cdot \vec \sigma) = 
\frac{1}{2} \left(\begin{array}{cc} 1 + e_3(x) & e_1(x) - i e_2(x) \\
e_1(x) + i e_2(x) & 1 - e_3(x) \end{array}\right) .
\end{equation}
In the above $CP(1)$ language, the lowest-order effective action of 
eq.(\ref{action}) takes the form
\begin{equation}
\label{Paction}
S[P] = \int d^2x \ dt \ \rho_s \mbox{Tr}\left[\p_i P \p_i P +
\frac{1}{c^2} \p_t P \p_t P\right].
\end{equation}
This action is invariant under the global transformations $g \in SU(2)_s$ of
eq.(\ref{trafog}),
\begin{equation}
P(x)' = g P(x) g^\dagger.
\end{equation}
Note that the magnon field $P(x)$ is invariant under the charge symmetries
$U(1)_Q$ and $SU(2)_Q$, i.e.\ $^{\vec Q}P(x) = P(x)$. 

\subsection{Discrete Symmetries of Magnon Fields}

Under the displacement $D$ by one lattice spacing the staggered magnetization 
chan\-ges sign, i.e.\ 
\begin{equation}
\label{shiftsym}
^D\vec e(x) = - \vec e(x) \ \Rightarrow \ ^DP(x) = \1 - P(x).
\end{equation}
Let us again combine $D$ with the spin rotation $g = i \sigma_2$, which results
in the transformation $D'$ with
\begin{equation}
^{D'}P(x) = (i \sigma_2)\ ^DP(x) (i \sigma_2)^\dagger = 
(i \sigma_2)[\1 - P(x)](i \sigma_2)^\dagger = P(x)^*,
\end{equation}
reminiscent of charge conjugation in particle physics.

The Hubbard model is invariant under translations by an integer multiple of the
lattice spacing. As we have seen, due to the antiferromagnetic order, the
displacement $D$ by one lattice spacing (which connects the two sublattices $A$
and $B$) plays a special role. In particular, in the effective theory it 
manifests itself as an internal symmetry that changes the sign of $\vec e(x)$. 
Translations by an even number of lattice spacings (which do not mix the 
sublattices), on the other hand, manifest themselves as ordinary translations 
in the effective theory. It should be noted that in the effective theory one 
need not distinguish between the displacement symmetries $D$ for the two 
spatial directions, since they are related by an ordinary translation by two 
lattice spacings (one in the $1$- and one in the $2$-direction). 

When we decompose a space-time vector $x = (x_1,x_2,t)$ into its spatial and 
temporal components, the 90 degrees rotation $O$ acts on $x$ as 
$Ox = (- x_2,x_1,t)$. Under the symmetry $O$ the magnon field transforms as
\begin{equation}
^OP(x) = P(Ox). 
\end{equation}
Similarly, under the spatial reflection $R$ at the $x_1$-axis, which turns $x$ 
into $Rx = (x_1,- x_2,t)$, the magnon field transforms as
\begin{equation}
^RP(x) = P(Rx).
\end{equation}
Had we not treated spin as an internal quantum number, it would also be 
directly affected by the spatial reflection. Since spin is a form of 
angular momentum, it transforms like the orbital angular momentum 
$\vec L = \vec r \times \vec p$ of a particle, which is a pseudo-vector and
thus changes into $^R\vec L = (- L_1,L_2,- L_3)$ under the reflection $R$.
This is equivalent to a 180 degrees $SU(2)_s$ rotation around the 
$2$-direction. Since we treat $SU(2)_s$ as an exact internal symmetry, the 
pure spatial inversion $R$ (without 180 degrees rotation of the spin) is also
a symmetry.

Another important symmetry is time-reversal $T$ which turns $x = (x_1,x_2,t)$ 
into $Tx = (x_1,x_2,- t)$. In a Hamiltonian description time-reversal is 
represented by an antiunitary operator. Here we discuss time-reversal in the 
framework of the Euclidean path integral. Again, the spin transforms like the
orbital angular momentum $\vec L$ of a particle. The momentum $\vec p$ changes 
sign under time-reversal and so does $\vec L$, i.e.\ 
$^T\vec L = - \vec L$.\footnote{Note that $- \vec L$ does not obey the angular 
momentum commutation relations. This is a consequence of the antiunitary 
nature of $T$ which does not represent an ordinary symmetry (implemented by a 
unitary transformation) in Hilbert space.} Consequently, under $T$ the 
staggered magnetization vector (which is built from microscopic spins) 
transforms as
\begin{equation}
^T\vec e(x) = - \vec e(Tx) \ \Rightarrow \
^TP(x) = \1 - P(Tx) = \ ^DP(Tx).
\end{equation}
Hence, time-reversal is closely related to the displacement symmetry of 
eq.(\ref{shiftsym}). Just like the displacement symmetry $D$, time-reversal is 
spontaneously broken in an antiferromagnet. However, in contrast to a 
ferromagnet, the combination $TD$ of time-reversal and the displacement 
symmetry remains unbroken. Previously we have combined the displacement 
symmetry $D$ with the $SU(2)_s$ spin rotation $i \sigma_2$ in order to obtain 
the unbroken symmetry $D'$. In order to obtain an unbroken variant $T'$ of 
time-reversal we now combine $T$ with the spin rotation $i \sigma_2$ which 
yields
\begin{equation}
^{T'}P(x) = (i \sigma_2) \ ^TP(x) (i \sigma_2)^\dagger =
(i \sigma_2) \ ^DP(Tx) (i \sigma_2)^\dagger = \ ^{D'}P(Tx).
\end{equation}

\subsection{Non-Linear Realization of the $SU(2)_s$ Symmetry}

In order to couple electron or hole fields to the magnons one must construct a 
non-linear realization of the spontaneously broken $SU(2)_s$ symmetry which 
then manifests itself as a local symmetry in the unbroken $U(1)_s$ subgroup of
$SU(2)_s$. This local transformation is constructed from the global 
transformation $g \in SU(2)_s$ as well as from the local magnon field $P(x)$ as
follows: one first diagonalizes the magnon field by a unitary transformation 
$u(x) \in SU(2)_s$, i.e.
\begin{equation}
u(x) P(x) u(x)^\dagger = \frac{1}{2}(\1 + \sigma_3) = 
\left(\begin{array}{cc} 1 & 0 \\ 0 & 0 \end{array} \right), \quad 
u_{11}(x) \geq 0.
\end{equation}
Note that, due to its projector properties, $P(x)$ has eigenvalues 0 and 1. In 
order to make $u(x)$ uniquely defined, we demand that the element $u_{11}(x)$ 
is real and non-negative. Otherwise the diagonalizing matrix $u(x)$ would be 
defined only up to a $U(1)_s$ phase. Using eq.(\ref{CP1O3}) and spherical 
coordinates for $\vec e(x)$, i.e.
\begin{equation}
\label{evec}
\vec e(x) = 
(\sin\theta(x) \cos\varphi(x),\sin\theta(x) \sin\varphi(x),\cos\theta(x)),
\end{equation}
one obtains
\begin{eqnarray}
u(x)&=&\frac{1}{\sqrt{2(1 + e_3(x))}}
\left(\begin{array}{cc} 1 + e_3(x) & e_1(x) - i e_2(x) \\ 
- e_1(x) - i e_2(x) & 1 + e_3(x) \end{array}\right) \nonumber \\
&=&\left(\begin{array}{cc} \cos(\frac{1}{2}\theta(x)) & 
\sin(\frac{1}{2}\theta(x)) \exp(- i \varphi(x)) \\
- \sin(\frac{1}{2}\theta(x)) \exp(i \varphi(x)) & \cos(\frac{1}{2}\theta(x)) 
\end{array}\right).
\end{eqnarray}
Under a global $SU(2)_s$ transformation $g$ the diagonalizing field $u(x)$
transforms as
\begin{equation}
\label{trafou}
u(x)' = h(x) u(x) g^\dagger, \quad u_{11}(x)' \geq 0,
\end{equation}
which implicitly defines the non-linear symmetry transformation 
\begin{equation}
\label{heq}
h(x) = \exp(i \alpha(x) \sigma_3) = \left(\begin{array}{cc} 
\exp(i \alpha(x)) & 0 \\ 0 & \exp(- i \alpha(x)) \end{array} \right) \in 
U(1)_s.
\end{equation}
The transformation $h(x)$ is uniquely defined since we demand that $u_{11}(x)'$
is again real and non-negative. Note that with this definition of $h(x)$ indeed
\begin{equation}
u(x)' P(x)' u(x)'^\dagger = \frac{1}{2}(\1 + \sigma_3).
\end{equation} 
Interestingly, the global $SU(2)_s$ transformation $g$ manifests itself in the 
form of a local transformation $h(x) \in U(1)_s$ which inherits its 
$x$-dependence from the magnon field $P(x)$.

We still need to show that the $SU(2)_s$ group structure $g = g_2 g_1$ is
inherited by the non-linear $U(1)_s$ realization, i.e.\ $h(x) = h_2(x) h_1(x)$. 
First, we perform the global $SU(2)_s$ transformation $g_1$, i.e.
\begin{equation}
P(x)' = g_1 P(x) g_1^\dagger, \quad u(x)' = h_1(x) u(x) g_1^\dagger,
\end{equation}
which defines the non-linear realization $h_1(x)$. Then we perform the 
subsequent global transformation $g_2$ which defines the non-linear realization
$h_2(x)$, i.e.
\begin{eqnarray}
&&P(x)'' = g_2 P(x)' g_2^\dagger = g_2 g_1 P(x) (g_2 g_1)^\dagger = 
g P(x) g^\dagger, \nonumber \\
&&u(x)'' = h_2(x) u(x)' g_2^\dagger = h_2(x) h_1(x) u(x) (g_2 g_1)^\dagger =
h(x) u(x) g^\dagger.
\end{eqnarray}
This indeed implies the correct group structure $h(x) = h_2(x) h_1(x)$.

Under the displacement symmetry $D$ the sign-change of the staggered 
magnetization $\vec e(x)$ implies
\begin{eqnarray}
\label{taueq}
\!\!\!\!\!\!\!\!\!\!^Du(x)&=&\frac{1}{\sqrt{2(1 - e_3(x))}}
\left(\begin{array}{cc} 1 - e_3(x) & - e_1(x) + i e_2(x) \\ 
e_1(x) + i e_2(x) & 1 - e_3(x) \end{array}\right) \nonumber \\
&=&\left(\begin{array}{cc} \sin(\frac{1}{2}\theta(x)) & 
- \cos(\frac{1}{2}\theta(x)) \exp(- i \varphi(x)) \\
\cos(\frac{1}{2}\theta(x)) \exp(i \varphi(x)) & \sin(\frac{1}{2}\theta(x)) 
\end{array}\right) \nonumber \\
&=&\tau(x) u(x),
\end{eqnarray}
where
\begin{eqnarray}
\tau(x)&=&\frac{1}{\sqrt{e_1(x)^2 + e_2(x)^2}}
\left(\begin{array}{cc} 0 & - e_1(x) + i e_2(x) \\
e_1(x) + i e_2(x) & 0 \end{array} \right) \nonumber \\
&=&\left(\begin{array}{cc} 0 & - \exp(- i \varphi(x)) \\
\exp(i \varphi(x)) & 0 \end{array} \right).
\end{eqnarray}
Note that $^D\tau(x) = - \tau(x) = \tau(x)^\dagger$, such that
\begin{equation}
^{DD}u(x) = \ ^D\tau(x) \ ^Du(x) = \tau(x)^\dagger \tau(x) u(x) = u(x),
\end{equation}
as one would expect for the displacement symmetry. It should also be noted that
--- like the $SU(2)_s$ symmetry --- the displacement symmetry is also 
spontaneously broken and hence realized in a non-linear (i.e.\ 
magnon-field-dependent) manner. Similarly, under the displacement symmetry $D'$
one finds $^{D'}u(x) = h(x)\ ^Du(x) g^\dagger$ with $g = i \sigma_2$. For this 
particular $g$ the local transformation takes the form 
$h(x) = (i \sigma_2) \tau(x)^\dagger$, such that
\begin{equation}
^{D'}u(x) = u(x)^*.
\end{equation}
In contrast to the displacement symmetry $D$, the symmetry $D'$ is not 
spontaneously broken and is thus realized in a linear (i.e.\ 
magnon-field-independent) manner.

In the next step we consider the anti-Hermitean field
\begin{equation}
\label{veq}
v_\mu(x) = u(x) \p_\mu u(x)^\dagger,
\end{equation} 
which transforms under $SU(2)_s$ as
\begin{equation}
\label{trafov}
v_\mu(x)' = h(x) u(x) g^\dagger \p_\mu [g u(x)^\dagger h(x)^\dagger] =
h(x) [v_\mu(x) + \p_\mu] h(x)^\dagger.
\end{equation}
Writing
\begin{equation}
v_\mu(x) = i v_\mu^a(x) \sigma_a = 
i \left(\begin{array}{cc} v_\mu^3(x) &  v_\mu^+(x) \\ v_\mu^-(x) & - v_\mu^3(x)
\end{array}\right), \quad
v_\mu^\pm(x) = v_\mu^1(x) \mp i v_\mu^2(x)
\end{equation}
and using eq.(\ref{heq}) this implies
\begin{equation}
v_\mu^3(x)' = v_\mu^3(x) - \p_\mu \alpha(x), \quad
v_\mu^\pm(x)' = \exp(\pm 2 i \alpha(x)) v_\mu^\pm(x).
\end{equation}
Hence, $v_\mu^3(x)$ transforms like an Abelian gauge field for $U(1)_s$, while
$v_\mu^\pm(x)$ represent vector fields ``charged'' under $U(1)_s$. For later
convenience we also introduce the Hermitean charged vector field
\begin{equation}
\label{Vfield}
V_\mu(x) = v_\mu^1(x) \sigma_1 + v_\mu^2(x) \sigma_2 = 
v_\mu^+(x) \sigma_+ + v_\mu^-(x) \sigma_- = 
\left(\begin{array}{cc} 0 &  v_\mu^+(x) \\ v_\mu^-(x) & 0 \end{array}\right),
\end{equation}
where $\sigma_\pm = \frac{1}{2}(\sigma_1 \pm i \sigma_2)$ are raising and
lowering operators of spin. Under the $SU(2)_s$ symmetry the charged vector 
field transforms as
\begin{equation}
V_\mu(x)' = h(x) V_\mu(x) h(x)^\dagger.
\end{equation}
The magnon action can also be written as
\begin{equation}
S[v_\mu] = \int d^2x \ dt \ 2 \rho_s 
\left(v_i^+ v_i^- + \frac{1}{c^2} v_t^+ v_t^-\right)
= \int d^2x \ dt \ \rho_s \mbox{Tr}\left[V_i^\dagger V_i + 
\frac{1}{c^2} V_t^\dagger V_t\right].
\end{equation}
It should be pointed out that the fields $v_\mu^a(x)$ do not represent 
independent degrees of freedom, but are composed of magnon fields. In 
particular, what looks like a mass term for a charged vector field is indeed
just the kinetic term of a massless Nambu-Goldstone boson.

\subsection{Discrete Symmetries of Composite Fields}

Under the displacement symmetry $D$ the composite vector field transforms as
\begin{eqnarray}
&&^Dv_\mu(x) = \tau(x)[v_\mu(x) + \p_\mu] \tau(x)^\dagger \ \Rightarrow \
^Dv_\mu^3(x) = - v_\mu^3(x) + \p_\mu \varphi(x), \nonumber \\
&&^Dv_\mu^\pm(x) = - \exp(\mp 2 i \varphi(x)) v_\mu^\mp(x), \quad
^DV_\mu(x) = \tau(x) V_\mu(x) \tau(x)^\dagger. 
\end{eqnarray}
Similarly, under the symmetry $D'$ one finds
\begin{eqnarray}
&&^{D'}v_\mu(x) = v_\mu(x)^* \ \Rightarrow \ ^{D'}v_\mu^3(x) = - v_\mu^3(x), 
\nonumber \\ 
&&^{D'}v_\mu^\pm(x) = - v_\mu^\mp(x), \quad ^{D'}V_\mu(x) = - V_\mu(x)^*.
\end{eqnarray}
This is exactly how an ordinary non-Abelian gauge field behaves under charge 
conjugation in particle physics.

Under the 90 degrees spatial rotation $O$ the composite field $v_\mu(x)$ 
transforms as
\begin{equation}
^Ov_i(x) = \varepsilon_{ij} v_j(Ox), \quad ^Ov_t(x) = v_t(Ox),
\end{equation}
while under the reflection $R$ one obtains
\begin{equation}
^Rv_1(x) = v_1(Rx), \quad ^Rv_2(x) = - v_2(Rx), \quad ^Rv_t(x) = v_t(Rx).
\end{equation}
Finally, under the time-reversal symmetry $T$ the field $v_\mu$ transforms as
\begin{eqnarray}
&&^Tv_i(x) = \ ^Dv_i(Tx), \quad ^Tv_t(x) = - \ ^Dv_t(Tx) \ \Rightarrow 
\nonumber \\
&&^Tv_i^3(x) = - v_i^3(Tx) + \p_i \varphi(Tx), \quad
^Tv_t^3(x) = v_t^3(Tx) - \p_t \varphi(Tx), \nonumber \\
&&^Tv_i^\pm(x) = - \exp(\mp 2 i \varphi(Tx)) v_i^\mp(Tx), \quad
^Tv_t^\pm(x) = \exp(\mp 2 i \varphi(Tx)) v_t^\mp(Tx), \nonumber \\
&&^TV_i(x) = \tau(Tx) V_i(Tx) \tau(Tx)^\dagger, \quad
^TV_t(x) = - \tau(Tx) V_t(Tx) \tau(Tx)^\dagger,
\end{eqnarray}
and under its unbroken variant $T'$ one finds
\begin{eqnarray}
&&^{T'}v_i(x) = \ ^{D'}v_i(Tx), \quad ^{T'}v_t(x) = - \ ^{D'}v_t(Tx), 
\nonumber \\
&&^{T'}v_i^3(x) = - v_i^3(Tx), \quad ^{T'}v_t^3(x) = v_t^3(Tx), \nonumber \\ 
&&^{T'}v_i^\pm(x) = - v_i^\mp(Tx), \quad ^{T'}v_t^\pm(x) = v_t^\mp(Tx), 
\nonumber \\ 
&&^{T'}V_i(x) = - V_i(Tx)^T, \quad ^{T'}V_t(x) = V_t(Tx)^T.
\end{eqnarray}
Note that the upper index $T$ on the right denotes transpose, while on the left
it denotes time-reversal. The above relations are equivalent to time-reversal 
of an ordinary non-Abelian gauge field.

\subsection{Alternative Representation of Magnon Fields}

We have used two equivalent representations of the magnon field in terms of the
unit-vector $\vec e(x)$ and in terms of the projection matrix $P(x)$. There is
a third equivalent representation in terms of a complex doublet
\begin{eqnarray}
\label{zspin}
&&z(x) = \left(\begin{array}{c} z_1(x) \\ z_2(x) \end{array}\right), \quad
z(x)^\dagger = (z_1(x)^*, z_2(x)^*), \nonumber \\
&&z(x)^\dagger z(x) = |z_1(x)|^2 + |z_2(x)|^2 = 1,
\end{eqnarray}
which is related to the other two representations by
\begin{eqnarray}
&&\vec e(x) = z(x)^\dagger \vec \sigma z(x) \ \Rightarrow \nonumber \\
&&e_1(x) = z_1(x)^* z_2(x) + z_2(x)^* z_1(x), \nonumber \\
&&e_2(x) = i [z_2(x)^* z_1(x) - z_1(x)^* z_2(x)], \nonumber \\
&&e_3(x) = |z_1(x)|^2 - |z_2(x)|^2, \nonumber \\
&&P(x) = z(x) z(x)^\dagger = 
\left(\begin{array}{cc} |z_1(x)|^2 & z_1(x) z_2(x)^* 
\\ z_2(x) z_1(x)^* & |z_2(x)|^2 \end{array}\right).
\end{eqnarray}
The field $z(x)$ is defined in terms of $\vec e(x)$ or $P(x)$ only up to a
$U(1)_s$ gauge transformation
\begin{equation}
\label{zgauge}
z(x)' = \exp(i \beta(x)) z(x).
\end{equation}
It is therefore necessary to also introduce the auxiliary real-valued $U(1)_s$ 
gauge field 
\begin{equation}
a_\mu(x) = \frac{1}{2 i}[z(x)^\dagger \p_\mu z(x) - \p_\mu z(x)^\dagger z(x)],
\end{equation}
which under the symmetry of eq.(\ref{zgauge}) transforms as
\begin{equation}
a_\mu(x)' = a_\mu(x) + \p_\mu \beta(x).
\end{equation}
The complex doublet $z(x)$ is closely related to the field $u(x)$. Fixing the
gauge freedom of eq.(\ref{zgauge}) such that $z_1(x)$ is real and non-negative,
it is easy to show that
\begin{equation}
u(x) = \left(\begin{array}{cc} z_1(x) & z_2(x)^* \\ - z_2(x) & z_1(x) 
\end{array}\right), \quad v_\mu^3(x) = a_\mu(x).
\end{equation}
Hence, the description in terms of complex doublets $z(x)$ and an additional
auxiliary gauge field $a_\mu(x)$ is physically equivalent to what we described
before. It should again be pointed out that $a_\mu(x)$ (or equivalently 
$v_\mu^3(x)$) does not represent a dynamical Abelian gauge field, but is 
simply a composite field constructed from the underlying magnon field $P(x)$.

\subsection{Baby-Skyrmions}

It is interesting to note that magnon fields support topological solitons
known as baby-Skyrmions --- a lower-dimensional variant of the Skyrme soliton
which represents a baryon in the low-energy pion effective theory for QCD
\cite{Sky61}. Baby-Skyrmions are solitons whose topological charge
\begin{equation}
B = \frac{1}{8 \pi} \int d^2x \ \varepsilon_{ij} 
\e \cdot (\p_i \e \times \p_j \e),
\end{equation}
defined at every instant in time, is an element of the homotopy group 
$\Pi_2[S^2] = \Z$. The corresponding topological current
\begin{equation}
\label{baby}
j_\mu(x) = \frac{1}{8 \pi} \varepsilon_{\mu\nu\rho} 
\e(x) \cdot [\p_\nu \e(x) \times \p_\rho \e(x)] 
\end{equation}
is conserved, i.e.\ $\p_\mu j_\mu = 0$, independent of the equations of motion.
Baby-Skyrmions are massive excitations inaccessible to the systematic 
low-energy expansion of chiral perturbation theory. Still, the existence of the
conserved current $j_\mu(x)$ may have physical consequences even for the pure 
magnon dynamics.

Under the various symmetries the topological current transforms as
\begin{eqnarray}
SU(2)_s:&&j_\mu(x)' = j_\mu(x), \nonumber \\
SU(2)_Q:&&^{\vec Q}j_\mu(x) = j_\mu(x), \nonumber \\
D:&&^Dj_\mu(x) = - j_\mu(x), \nonumber \\
D':&&^{D'}j_\mu(x) = - j_\mu(x), \nonumber \\
O:&&^Oj_t(x) = j_t(Ox), \quad ^Oj_i(x) = \varepsilon_{ij} j_j(Ox), \nonumber \\
R:&&^Rj_t(x) = - j_t(Rx), \quad ^Rj_1(x) = - j_1(Rx), \quad ^Rj_2(x) = j_2(Rx),
\nonumber \\
T:&&^Tj_t(x) = - j_t(Tx), \quad ^Tj_1(x) = j_1(Tx), \quad ^Tj_2(x) = j_2(Rx), 
\nonumber \\
T':&&^{T'}j_t(x) = - j_t(Tx), \quad ^{T'}j_1(x) = j_1(Tx), \quad 
^{T'}j_2(x) = j_2(Tx).
\end{eqnarray}
One might be tempted to add a term $j_\mu(x) v_\mu^3(x)$ to the magnon 
Lagrangian because this is how an Abelian gauge field couples to a conserved 
current. Indeed, this term is invariant under $SU(2)_s$, $SU(2)_Q$, $D$, $D'$, 
and $O$. However, it violates the reflection and time-reversal symmetries $R$, 
$T$, and $T'$ and is hence forbidden. 

There is another non-trivial homotopy group, $\Pi_3[S^2] = \Z$, which is 
relevant for baby-Skyrmions. It implies that space-time-dependent magnon fields
fall into distinct topological classes characterized by the Hopf number 
$H[\e] \in \Pi_3[S^2] = \Z$. In $2+1$ dimensions baby-Skyrmions can be 
quantized as anyons characterized by a statistics angle $\theta$ \cite{Wil83}. 
The cases $\theta = 0$ and $\theta = \pi$ correspond to bosons and fermions, 
respectively. Including the Hopf term, the magnon path integral takes the form
\begin{equation}
\label{magnonPI}
Z = \int {\cal D}\e \ \exp(- S[\e]) \exp(i \theta H[\e]).
\end{equation}
The Hopf term also changes sign under $R$, $T$, and $T'$. Hence, 
$\exp(i \theta H[\e])$ is invariant only if $\theta$ is 0 or $\pi$. 
Consequently, in an antiferromagnet with exact $R$, $T$, or $T'$ symmetries 
baby-Skyrmions can only be quantized as bosons or fermions. For the 
antiferromagnetic quantum Heisenberg model it has been argued that no Hopf 
term is generated \cite{Wen88,Hal88,Dom88,Fra88,Rea89}. Hence, in that case the
baby-Skyrmions should be bosons.

\section{The Hubbard Model in a Magnon Background Field}

The half-filled ground state of the Hubbard model plays a similar role as the 
Dirac sea in a relativistic quantum field theory. In particular, any fermion 
added to a half-filled state will be denoted as an electron, while any fermion 
removed from such a state represents a hole. In this section we couple a 
background magnon field to the microscopic degrees of freedom of the Hubbard 
model. In this way composite operators are constructed which transform exactly
like the fields of the effective theory. Hence, the effective fields 
inherit their transformation properties under symmetry operations from the 
Hubbard model degrees of freedom.

\subsection{Fermion Operators in a Magnon Background Field}

In order to analyze the transformation properties of the electron and hole 
fields, as an intermediate step between the microscopic and effective 
descriptions, we first add a continuum magnon background field $P(x)$ to the 
Hubbard model by hand. The corresponding diagonalizing unitary matrix field 
$u(x)$ is used to turn the matrix-valued Hubbard model operator $C_x$ of 
eq.(\ref{Coperator}) into new operators $\Psi^A_x$ and $\Psi^B_x$ defined on 
the even and odd sublattices, respectively
\begin{eqnarray}
&&\Psi^A_x = u(x) C_x = u(x) \left(\begin{array}{cc} c_{x\uparrow} & 
c_{x \downarrow}^\dagger \\ c_{x \downarrow} & 
- c_{x\uparrow}^\dagger\end{array} \right) = 
\left(\begin{array}{cc} \psi^A_{x+} & \psi^{A\dagger}_{x-} \\ 
\psi^A_{x-} & - \psi^{A\dagger}_{x+} \end{array} \right), \quad x \in A, 
\nonumber \\
&&\Psi^B_x = u(x) C_x = u(x) \left(\begin{array}{cc} c_{x\uparrow} 
&- c_{x \downarrow}^\dagger \\ c_{x \downarrow} & c_{x\uparrow}^\dagger 
\end{array} \right) = 
\left(\begin{array}{cc} \psi^B_{x+} & - \psi^{B\dagger}_{x-} \\ 
\psi^B_{x-} & \psi^{B\dagger}_{x+} \end{array} \right), \quad x \in B.
\end{eqnarray}
In order to achieve a consistent representation of the underlying 
antiferromagnetic structure, it is unavoidable to explicitly split the degrees 
of freedom according to their location on sublattice $A$ or $B$. In this 
context it may be interesting to consider the electron-hole representation of 
the Hubbard model operators discussed in appendix A. The operators 
$\psi^{A,B}_{x \pm}$ obey standard anticommutation relations. It should be 
noted that here the continuum field $u(x)$ is evaluated only at discrete 
lattice points $x$.

The new lattice operators inherit their transformation properties from the
operators of the Hubbard model. According to eqs.(\ref{trafou}) and 
(\ref{trafog}), under the $SU(2)_s$ symmetry one obtains
\begin{equation}
{\Psi_x^{A,B}}' = u(x)' C_x' = h(x) u(x) g^\dagger g C_x = h(x) \Psi_x^{A,B}. 
\end{equation}
In components this relation takes the form
\begin{equation}
\label{trafopsi}
{\psi^{A,B}_{x\pm}}' = \exp(\pm i \alpha(x)) \psi^{A,B}_{x\pm}.
\end{equation}
The components $\psi^{A,B}_{x\pm}$ do not simply correspond to spin up and spin
down with respect to an arbitrarily chosen global quantization axis. Instead 
they correspond to spin parallel ($+$) or antiparallel ($-$) to the local 
staggered magnetization. This follows from considering global symmetry 
transformations $g \in U(1)_s$ in the unbroken subgroup of $SU(2)_s$ which 
describe rotations around the spontaneously selected direction of the staggered
magnetization vector. In that case, according to eq.(\ref{trafou}), $h(x) = g$ 
becomes a global transformation as well and eq.(\ref{trafopsi}) shows that 
$\psi^{A,B}_{x\pm}$ indeed has spin parallel or antiparallel to the direction 
of the staggered magnetization. 

Similarly, under the $SU(2)_Q$ symmetry one obtains
\begin{equation}
^{\vec Q}\Psi_x^{A,B} = \ ^{\vec Q}u(x) ^{\vec Q}C_x = u(x) C_x \Omega^T = 
\Psi_x^{A,B} \Omega^T. 
\end{equation}
In particular, under the $U(1)_Q$ subgroup of $SU(2)_Q$ the components 
transform as
\begin{equation}
^Q\psi^{A,B}_{x\pm} = \exp(i \omega) \psi^{A,B}_{x\pm}.
\end{equation}
Under the displacement symmetry the new operators transform as
\begin{equation}
^D\Psi^{A,B}_x = \ ^Du(x+\hat i) C_{x+\hat i} \sigma_3 = 
\tau(x+\hat i) u(x+\hat i) C_{x+\hat i} \sigma_3 = 
\tau(x+\hat i) \Psi^{B,A}_{x+\hat i} \sigma_3,
\end{equation}
where $\tau(x)$ is the field introduced in eq.(\ref{taueq}). Expressed in
components this implies
\begin{equation}
^D\psi^{A,B}_{x\pm} = \mp \exp(\mp i \varphi(x+\hat i)) 
\psi^{B,A}_{x+\hat i,\mp}.
\end{equation}
Similarly, under the symmetry $D'$ one finds
\begin{equation}
^{D'}\Psi^{A,B}_x = \ ^{D'}u(x+\hat i) (i \sigma_2) C_{x+\hat i} \sigma_3 = 
u(x+\hat i)^* (i \sigma_2) C_{x+\hat i} \sigma_3 = 
(i \sigma_2) \Psi^{B,A}_{x+\hat i} \sigma_3.
\end{equation}
Here we have used $u(x+\hat i)^* (i \sigma_2) = (i \sigma_2) u(x+\hat i)$.
Again, expressed in components this relation takes the form
\begin{equation}
^{D'}\psi^{A,B}_{x\pm} = \pm \psi^{B,A}_{x+\hat i,\mp}.
\end{equation}
We have seen before that the symmetry $D'$ acts on the composite field 
$v_\mu(x)$ exactly like charge conjugation in particle physics. However, it 
should be noted that $D'$ acts on the electron and hole fields in a different 
way than the usual charge conjugation of a relativistic Dirac fermion which 
interchanges electrons and positrons. In particular, $D'$ does not interchange 
electrons and holes. Instead, it flips the spin of both electrons and holes 
from $+$ to $-$ and vice versa. Indeed, the spin is the ``charge'' that couples
to the composite gauge field of eq.(\ref{veq}) constructed from the magnon 
field.

In the condensed matter literature on high-temperature superconductivity the
concept of spin-charge separation (whose existence is established for some 
systems in one spatial dimension) has often been invoked. The idea is that 
there may be quasi-particles --- so-called holons --- which carry charge but 
no spin, as well as so-called spinons which are neutral and carry spin 
$\frac{1}{2}$. In order to avoid confusion between holons and the holes of our 
effective theory, we like to make a few comments: one might think that the
fermion operator $\Psi^{A,B}_x$ does not carry spin since it does not transform
with the global spin transformation $g \in SU(2)_s$. However, the spin symmetry
is non-linearly realized and hence the fermion operator transforms with the 
local $h(x) \in U(1)_s$. Consequently, $\Psi^{A,B}_x$ still carries spin and 
hence does not represent a holon. It should also be pointed out that in the 
weakly coupled effective theory of magnons and holes there are no linearly 
confining forces that could form a spinless holon out of $\Psi^{A,B}_x$ and the
magnon field $z(x)$ of eq.(\ref{zspin}).

\subsection{Formal Continuum Limit of the Hubbard Model in a Magnon Background 
Field}

In terms of the new operators the Hubbard model Hamiltonian takes the form
\begin{eqnarray}
\label{Hpsi}
H&=&- \frac{t}{2} \sum_{x \in A,i} 
\mbox{Tr}[\Psi_x^{A\dagger} {\cal V}_{x,i} \Psi^B_{x+\hat i} +
\Psi_{x+\hat i}^{B\dagger} {\cal V}_{x,i}^\dagger \Psi^A_x] \nonumber \\
&&- \frac{t}{2} \sum_{x \in B,i} 
\mbox{Tr}[\Psi_x^{B\dagger} {\cal V}_{x,i} \Psi^A_{x+\hat i} +
\Psi_{x+\hat i}^{A\dagger} {\cal V}_{x,i}^\dagger \Psi^B_x] \nonumber \\
&&+ \frac{U}{12} \sum_{x \in A} 
\mbox{Tr}[\Psi_x^{A\dagger} \Psi^A_x \Psi_x^{A\dagger} \Psi^A_x]
+ \frac{U}{12} \sum_{x \in B} 
\mbox{Tr}[\Psi_x^{B\dagger} \Psi^B_x \Psi_x^{B\dagger} \Psi^B_x] \nonumber \\
&&- \frac{\mu}{2} \sum_{x \in A} \mbox{Tr}[\Psi_x^{A\dagger} \Psi^A_x \sigma_3]
- \frac{\mu}{2} \sum_{x \in B} \mbox{Tr}[\Psi_x^{B\dagger} \Psi^B_x \sigma_3],
\end{eqnarray}
where we have introduced the parallel transporter
\begin{equation}
{\cal V}_{x,i} = u(x) u(x+\hat i)^\dagger \in SU(2)_s, 
\end{equation}
which transforms under $SU(2)_s$ as
\begin{equation}
{\cal V}_{x,i}' = h(x) {\cal V}_{x,i} h(x+\hat i)^\dagger.
\end{equation}
For smooth magnon fields we can put
\begin{eqnarray}
&&u(x) = u(x+\frac{\hat i}{2}) - \frac{a}{2} \p_i u(x+\frac{\hat i}{2}) +
\frac{a^2}{8} \p_i^2 u(x+\frac{\hat i}{2}) + {\cal O}(a^3), \nonumber \\
&&u(x+\hat i) = u(x+\frac{\hat i}{2}) + 
\frac{a}{2} \p_i u(x+\frac{\hat i}{2}) + 
\frac{a^2}{8} \p_i^2 u(x+\frac{\hat i}{2}) + {\cal O}(a^3),
\end{eqnarray}
where $a$ is the lattice spacing. Similar expressions hold for the other 
fields. Using the unitarity of $u(x+\frac{\hat i}{2})$ one can show that the 
lattice parallel transporter reduces to
\begin{equation}
{\cal V}_{x,i} = \1 + a v_i(x+\frac{\hat i}{2}) + 
\frac{a^2}{2} v_i(x+\frac{\hat i}{2})^2 + {\cal O}(a^3),
\end{equation}
with $v_i(x)$ given by eq.(\ref{veq}). Note that both the continuum field 
$v_i(x)$ and the lattice parallel transporter field ${\cal V}_{x,i}$ transform 
locally only with the unbroken $U(1)_s$ subgroup and not with the full 
$SU(2)_s$ symmetry.

In the continuum limit we make the replacements
\begin{equation}
\sum_{x \in A}, \ \sum_{x \in B} \ \rightarrow \ \frac{1}{2 a^2} \int d^2x, 
\quad \Psi^{A,B}_x \ \rightarrow \ \sqrt{2} a \Psi^{A,B}(x).
\end{equation}
The factor $\frac{1}{2}$ in front of the integral accounts for the fact that
each sublattice covers only half of the space. Similarly the factor 
$\sqrt{2} a$ in the definition of the continuum field $\Psi^{A,B}(x)$ arises 
because there is only one degree of freedom of a given type $A$ or $B$ per area
$2 a^2$. The components $\psi_\pm^{A,B}(x)$ of $\Psi^{A,B}(x)$ again obey 
standard anticommutation relations, however, with the Dirac $\delta$-function 
of the continuum theory instead of the Kronecker $\delta$-function of the 
lattice. It should be noted that, due to the antiferromagnetic order, the 
number of degrees of freedom per continuum point is twice as large as the 
number per lattice point. Taking the formal continuum limit $a \rightarrow 0$ 
(and ignoring an irrelevant constant) the Hamiltonian of eq.(\ref{Hpsi}) takes 
the form
\begin{eqnarray}
H&=&\int d^2x \ 
\{M \mbox{Tr}[\Psi^{A\dagger} \Psi^B] + 
\frac{1}{2 M'} \mbox{Tr}[D_i \Psi^{A\dagger} D_i \Psi^B] \nonumber \\
&&+ i K \mbox{Tr}[D_i \Psi^{A\dagger} V_i \Psi^B +
D_i \Psi^{B\dagger} V_i \Psi^A] + N \mbox{Tr}[\Psi^{A\dagger} V_i V_i \Psi^B] 
\nonumber \\
&&+ \frac{G}{12} \mbox{Tr}[\Psi^{A\dagger} \Psi^A \Psi^{A\dagger} \Psi^A +
\Psi^{B\dagger} \Psi^B \Psi^{B\dagger} \Psi^B] -
\frac{\mu}{2} \mbox{Tr}[\Psi^{A\dagger} \Psi^A \sigma_3 +
\Psi^{B\dagger} \Psi^B \sigma_3]\}. \nonumber \\ \ 
\end{eqnarray}
It should be noted that, due to the structure of $\Psi^{A,B}(x)$, the 
individual terms are Hermitean. In the above expression $V_\mu(x)$ is the field
defined in eq.(\ref{Vfield}) and the covariant derivatives are given by
\begin{eqnarray}
\label{covarder}
&&D_\mu \Psi^{A,B}(x) = (\p_\mu + i v_\mu^3(x) \sigma_3) \Psi^{A,B}(x), 
\nonumber \\
&&D_\mu \Psi^{A,B\dagger}(x) = [D_\mu \Psi^{A,B}(x)]^\dagger =
\p_\mu \Psi^{A,B\dagger}(x) - \Psi^{A,B\dagger}(x) i v_\mu^3(x) \sigma_3.
\end{eqnarray}
In terms of the fundamental parameters $t$ and $U$ and the lattice spacing $a$ 
of the Hubbard model one obtains
\begin{equation}
M = - 4 t, \quad M' = \frac{1}{2 t a^2}, \quad K = t a^2, \quad N = t a^2, 
\quad G = 2 U a^2.
\end{equation}
It should be noted that (in contrast to a relativistic theory) the kinetic mass
$M'$ is in general different from the rest mass $M$. The Hamiltonian from above
resembles some (but not all) terms in the action of the effective theory to be 
constructed below. However, the coupling constants resulting from the formal 
continuum limit get renormalized and will hence be replaced by a priori unknown
low-energy parameters in the effective action. The values of the low-energy 
parameters can be determined in experiments with cuprate materials or through 
numerical simulations of a microscopic Hubbard-type model.

\section{Effective Theory for Magnons and Charge Carriers}

The low-energy effective theory for magnons is analogous to chiral perturbation
theory for pions in QCD. In QCD the baryon number $B$ is a conserved quantity. 
Thus one can investigate the low-energy QCD dynamics separately in each baryon 
number sector. Ordinary chiral perturbation theory operates in the $B = 0$ 
sector. The low-energy physics in the $B = 1$ sector involves a single nucleon 
interacting with soft pions. The low-energy effective theory describing these 
dynamics is known as baryon chiral perturbation theory 
\cite{Geo84a,Gas88,Jen91,Ber92,Bec99}. Similar effective theories have been 
constructed for the $B = 2$ \cite{Wei90,Kap98} and $B = 3$ sectors 
\cite{Bed98,Bed02} in the context of nuclear physics. Even nuclear matter 
(i.e.\ a system with non-zero baryon density) has been studied with effective 
theories \cite{Rho00,Par00,Mue00,Mei02,Wir03}.
The condensed matter analog of baryon number is electron (or hole) number (or 
equivalently electric charge) which is obviously also conserved. In analogy to
QCD it is hence possible to construct a low-energy effective theory describing 
the interactions of soft magnons with charge carriers doped into an 
antiferromagnet. Most high-$T_c$ materials result by hole-doping of quantum 
antiferromagnets, but the effective theory also applies to electron-doping. The
key observation is that the spontaneously broken $SU(2)_s$ spin symmetry is 
non-linearly realized on the electron or hole fields and appears as a local 
$U(1)_s$ symmetry in the unbroken subgroup. This is analogous to baryon chiral 
perturbation theory in which the spontaneously broken $SU(2)_L \otimes SU(2)_R$
chiral symmetry of QCD is implemented on the nucleon fields as a local 
$SU(2)_{L=R}$ transformation in the unbroken isospin subgroup.

\subsection{Effective Fields for Charge Carriers} 

In the low-energy effective theory we will use a Euclidean path integral
description instead of the Hamiltonian description used in the Hubbard model.
Consequently, the Hermitean conjugate lattice operators 
$\psi^{A,B\dagger}_{x\pm}$ are then replaced by Grassmann numbers 
$\psi^{A,B\dagger}_\pm(x)$ which are completely independent of 
$\psi^{A,B}_\pm(x)$. Therefore, in the effective theory the electron and hole 
fields are represented by eight independent Grassmann numbers 
$\psi^{A,B}_\pm(x)$ and $\psi^{A,B\dagger}_\pm(x)$ which can be combined to
\begin{equation}
\label{phi}
\Psi^A(x) = \left(\begin{array}{cc} \psi^A_+(x) & \psi^{A\dagger}_-(x) \\ 
\psi^A_-(x) & - \psi^{A\dagger}_+(x) \end{array} \right), \quad
\Psi^B(x) = \left(\begin{array}{cc} \psi^B_+(x) & - \psi^{B\dagger}_-(x) \\ 
\psi^B_-(x) & \psi^{B\dagger}_+(x) \end{array} \right).
\end{equation}
In order to avoid confusion with relativistic theories, we do not denote the 
conjugate fields by $\overline{\psi}^{A,B}_\pm(x)$. For notational convenience 
we also introduce the fields
\begin{equation}
\label{phidagger}
\Psi^{A\dagger}(x) = \left(\begin{array}{cc} \psi^{A\dagger}_+(x) & 
\psi^{A\dagger}_-(x) \\ \psi^A_-(x) & - \psi^A_+(x) \end{array} \right), \quad
\Psi^{B\dagger}(x) = \left(\begin{array}{cc} \psi^{B\dagger}_+(x) & 
\psi^{B\dagger}_-(x) \\ - \psi^B_-(x) & \psi^B_+(x) \end{array} \right).
\end{equation}
It should be noted that $\Psi^{A,B\dagger}(x)$ is not an independent field, but
consists of the same Grassmann fields $\psi^{A,B}_\pm(x)$ and 
$\psi^{A,B\dagger}_\pm(x)$ as $\Psi^{A,B}(x)$.

It should be pointed out that, since they emerge dynamically, the continuum 
fields of the low-energy effective theory can not be derived explicitly from 
the lattice operators of the microscopic Hubbard model. Still, the Grassmann 
fields $\Psi^{A,B}(x)$ describing electrons and holes in the low-energy 
effective theory
transform just like the lattice operators $\Psi^{A,B}_x$ discussed before. In 
contrast to the lattice operators, the fields $\Psi^{A,B}(x)$ are defined in 
the continuum. Hence, under the displacement symmetries $D$ and $D'$ one no 
longer distinguishes between the points $x$ and $x+\hat i$. As a result, the
transformation rules of the various symmetries take the form
\begin{eqnarray}
\label{Psitrafo}
SU(2)_s:&&\Psi^{A,B}(x)' = h(x) \Psi^{A,B}(x), \nonumber \\
SU(2)_Q:&&^{\vec Q}\Psi^{A,B}(x) = \Psi^{A,B}(x) \Omega^T, \nonumber \\
D:&&^D\Psi^{A,B}(x) = \tau(x) \Psi^{B,A}(x) \sigma_3, \nonumber \\
D':&&^{D'}\Psi^{A,B}(x) = (i \sigma_2) \Psi^{B,A}(x) \sigma_3.
\end{eqnarray}
In components the symmetry transformations read
\begin{eqnarray}
\label{psitrafo1}
SU(2)_s:&&\psi^{A,B}_\pm(x)' = \exp(\pm i \alpha(x)) \psi^{A,B}_\pm(x),
\nonumber \\
U(1)_Q:&&^Q\psi^{A,B}_\pm(x) = \exp(i \omega) \psi^{A,B}_\pm(x),
\nonumber \\
D:&&^D\psi^{A,B}_\pm(x) = \mp \exp(\mp i \varphi(x)) \psi^{B,A}_\mp(x),
\nonumber \\
D':&&^{D'}\psi^{A,B}_\pm(x) = \pm \psi^{B,A}_\mp(x).
\end{eqnarray}
Under the space-time symmetries, i.e.\ under the 90 degrees rotation $O$, the
reflection $R$, time-reversal $T$, and its unbroken variant $T'$ the fermion 
fields transform as
\begin{eqnarray}
O:&&^O\Psi^{A,B}(x) = \Psi^{A,B}(Ox), \nonumber \\
R:&&^R\Psi^{A,B}(x) = \Psi^{A,B}(Rx), \nonumber \\
T:&&^T\Psi^{A,B}(x) = \tau(Tx) (i \sigma_2) \Psi^{A,B\dagger}(Tx)^T \sigma_3, 
\nonumber \\
&&^T\Psi^{A,B\dagger}(x) = - \sigma_3 \Psi^{A,B}(Tx)^T (i \sigma_2)^\dagger
\tau(Tx)^\dagger, 
\nonumber \\
T':&&^{T'}\Psi^{A,B}(x) = - \Psi^{A,B\dagger}(Tx)^T \sigma_3, \nonumber \\
&&^{T'}\Psi^{A,B\dagger}(x) = \sigma_3 \Psi^{A,B}(Tx)^T.
\end{eqnarray}
Again an upper index $T$ on the right denotes transpose, while on the left it 
denotes time-reversal. The form of the time-reversal symmetry $T$ in the 
effective theory with non-linearly realized $SU(2)_s$ symmetry follows from the
usual form of time-reversal in the Euclidean path integral of a 
non-relativistic theory in which the spin symmetry is linearly realized. The 
fermion fields in the two formulations just differ by a factor $u(x)$. In 
components the previous relations take the form
\begin{eqnarray}
\label{psitrafo2}
O:&&^O\psi^{A,B}_\pm(x) = \psi^{A,B}_\pm(Ox), \nonumber \\
R:&&^R\psi^{A,B}_\pm(x) = \psi^{A,B}_\pm(Rx), \nonumber \\
T:&&^T\psi^{A,B}_\pm(x) =  \exp(\mp i \varphi(Tx)) \psi^{A,B\dagger}_\pm(Tx),
\nonumber \\
&&^T\psi^{A,B\dagger}_\pm(x) = - \exp(\pm i \varphi(Tx)) \psi^{A,B}_\pm(Tx), 
\nonumber \\
T':&&^{T'}\psi^{A,B}_\pm(x) = - \psi^{A,B\dagger}_\pm(Tx), \nonumber \\
&&^{T'}\psi^{A,B\dagger}_\pm(x) = \psi^{A,B}_\pm(Tx).
\end{eqnarray}
It should be noted that the components $+$ and $-$ (denoting spin parallel and
antiparallel to the direction of the staggered magnetization) are not
interchanged under time-reversal. While both the spin of the fermion and the
staggered magnetization change sign under time-reversal, the projection of
one onto the other does not. 

The action to be constructed in the next section must be invariant under the 
internal symmetries $SU(2)_s$, $U(1)_Q$ (or even $SU(2)_Q$), $D$ and $D'$, as 
well as under space-time translations and the other space-time symmetries $O$, 
$R$, and $T$ (or equivalently $T'$).

The fundamental forces underlying condensed matter physics are 
Poincar\'e-in\-va\-ri\-ant. However, some of the space-time symmetries may be 
spontaneously broken by the formation of a crystal lattice. The resulting
Nambu-Goldstone bosons are the phonons, which play a central role in ordinary 
low-temperature superconductors by providing the force that binds Cooper pairs.
In high-$T_c$ superconductors, on the other hand, it is expected that phonons 
alone cannot provide the mechanism for Cooper pair formation. In the Hubbard 
model (and also in our effective theory) phonons are explicitly excluded 
because one imposes a rigid lattice by hand. This does not only break 
continuous translations and rotations down to their discrete counterparts; it 
also breaks space-time rotations. In a relativistic context these would be the 
boosts of the Poincar\'e group. In a non-relativistic theory the lattice 
explicitly breaks Galilean boost invariance, thus providing a preferred rest 
frame (a condensed matter ``ether''). As a consequence, the magnon-mediated 
forces between a pair of electrons or holes may depend on the center of mass 
momentum of the pair. In the actual high-$T_c$ materials Galilean (or more
precisely Poincar\'e symmetry) is spontaneously (and not explicitly) broken. If
phonons play an important role in the understanding of high-temperature 
superconductivity, one should construct an effective theory of spontaneously 
broken (and thus non-linearly realized) $SU(2)_s$ and Galilean symmetry which 
would automatically include both magnons and phonons. This is indeed possible 
and presently under investigation using the techniques of low-energy effective 
field theory. In the present paper we assume that phonons play no major role in
the cuprates. In that case, it is legitimate to break Galilean invariance 
explicitly instead of spontaneously.

\subsection{Effective Action for Magnons and Charge Carriers}

We now construct the leading terms in the effective action of magnons and 
electrons or holes. The effective theory provides a systematic low-energy 
expansion organized according to the number of derivatives in the terms of the 
effective action. We decompose the effective Lagrangian into an 
$SU(2)_Q$-invariant part ${\cal L}$ and an $SU(2)_Q$-breaking (but still 
$U(1)_Q$-invariant) part $\widetilde{\cal L}$. The contributions 
${\cal L}_{n_t,n_i,n_\psi}$ and $\widetilde{\cal L}_{n_t,n_i,n_\psi}$ to the 
effective Lagrangian are classified according to the number of time-derivatives
$n_t$, the number of spatial derivatives $n_i$, and the number of fermion 
fields $n_\psi$ they contain. The total action is then given by
\begin{equation}
S[\psi^{A,B\dagger}_\pm,\psi^{A,B}_\pm,P] = 
\int d^2x \ dt \ \sum_{n_t,n_i,n_\psi}
({\cal L}_{n_t,n_i,n_\psi} + \widetilde {\cal L}_{n_t,n_i,n_\psi})
\end{equation}
and the partition function takes the form
\begin{equation}
Z = \int {\cal D}\psi^{A,B\dagger}_\pm {\cal D}\psi^{A,B}_\pm {\cal D}P
\exp(- S[\psi^{A,B\dagger}_\pm,\psi^{A,B}_\pm,P]).
\end{equation}

Until now we have constructed the effective action in the $Q = 0$ sector, i.e.\
for a half-filled system which is described entirely in terms of magnons. Since
antiferromagnetic magnons have a ``relativistic'' dispersion relation (with the
spin-wave velocity $c$ playing the role of the velocity of light), in pure
magnon chiral perturbation theory one counts temporal and spatial derivatives 
as being of the same order. The leading contributions of eq.(\ref{Paction}) 
take the form
\begin{equation}
{\cal L}_{2,0,0} = \frac{\rho_s}{c^2} \mbox{Tr}[\p_t P \p_t P], \quad
{\cal L}_{0,2,0} = \rho_s \mbox{Tr}[\p_i P \p_i P].
\end{equation}

Next we consider terms quadratic in the fermion fields. These contribute to the
scattering of magnons off electrons or holes in the $Q = \pm 1$ sectors and 
they generally describe the propagation of charge carriers in an 
antiferromagnet with $|Q| \geq 1$. In contrast to magnons, electrons or holes 
are massive and have a non-relativistic dispersion relation. Hence, it is
natural to count one temporal and two spatial derivatives as being of the same 
order. In order to count derivatives consistently, in the $Q \neq 0$ sectors it
may thus be necessary to also consider the pure magnon term ${\cal L}_{2,0,0}$ 
with two temporal derivatives as being of higher order. The leading order terms
without any derivatives which are Hermitean and invariant under $SU(2)_s$, 
$SU(2)_Q$, $D$, and $D'$ as well as under the space-time symmetries $O$, $R$, 
$T$, and $T'$ take the form
\begin{eqnarray}
{\cal L}_{0,0,2}&=& 
M_1 \mbox{Tr}[\Psi^{A\dagger} \Psi^B] 
+ \frac{M_2}{2} \mbox{Tr}[\Psi^{A\dagger} \sigma_3 \Psi^A - 
\Psi^{B\dagger} \sigma_3 \Psi^B] \nonumber \\
&=&M_1 (\psi^{A\dagger}_+ \psi^B_+ + \psi^{A\dagger}_- \psi^B_- +
\psi^{B\dagger}_+ \psi^A_+ + \psi^{B\dagger}_- \psi^A_-) \nonumber \\
&&+ M_2 (\psi^{A\dagger}_+ \psi^A_+ - \psi^{A\dagger}_- \psi^A_-
- \psi^{B\dagger}_+ \psi^B_+ + \psi^{B\dagger}_- \psi^B_-).
\end{eqnarray}
The mass parameters $M_1$ and $M_2$ (as well as all other low-energy parameters
to be introduced below) take real values in order to ensure Hermiticity of the
corresponding Hamiltonian. It should be noted that
\begin{eqnarray}
&&\mbox{Tr}[\Psi^{A\dagger} \Psi^A] = \mbox{Tr}[\Psi^{B\dagger} \Psi^B] = 0,
\nonumber \\ 
&&\mbox{Tr}[\Psi^{A\dagger} \Psi^B] = \mbox{Tr}[\Psi^{B\dagger} \Psi^A],
\end{eqnarray}
due to the anticommutativity of the Grassmann fields. When we impose only the 
generic $U(1)_Q$ but not the full $SU(2)_Q$ symmetry, one more fermion mass 
term can be added
\begin{eqnarray}
\widetilde{\cal L}_{0,0,2}&=&\frac{m}{2}
\mbox{Tr}[\Psi^{A\dagger} \Psi^A \sigma_3 + \Psi^{B\dagger} \Psi^B \sigma_3]
\nonumber \\
&=&m (\psi^{A\dagger}_+ \psi^A_+ + \psi^{A\dagger}_- \psi^A_- + 
\psi^{B\dagger}_+ \psi^B_+ + \psi^{B\dagger}_- \psi^B_-).
\end{eqnarray}
This term can be absorbed into a redefinition of the chemical potential. 
Remarkably, no other fermion mass terms (consistent with the $SU(2)_s$, 
$U(1)_Q$, $D$, $D'$, $T$, and $T'$ symmetries) exist. In particular, it is
useful to note that
\begin{equation}
\mbox{Tr}[\Psi^{A\dagger} \sigma_3 \Psi^A \sigma_3] = 
\mbox{Tr}[\Psi^{B\dagger} \sigma_3 \Psi^B \sigma_3] = 0.
\end{equation}

The terms with one temporal derivative are given by
\begin{eqnarray}
{\cal L}_{1,0,2}&=&\frac{1}{2} \mbox{Tr}[\Psi^{A\dagger} D_t \Psi^A +
\Psi^{B\dagger} D_t \Psi^B] \nonumber \\
&&+ \frac{\Lambda_1}{2} \mbox{Tr}[\Psi^{A\dagger} V_t \Psi^A + 
\Psi^{B\dagger} V_t \Psi^B] + 
\Lambda_2 \mbox{Tr}[\Psi^{A\dagger} \sigma_3 V_t \Psi^B] \nonumber \\
&=&\psi^{A\dagger}_+ D_t \psi^A_+ + \psi^{A\dagger}_- D_t \psi^A_- 
+ \psi^{B\dagger}_+ D_t \psi^B_+ + \psi^{B\dagger}_- D_t \psi^B_- \nonumber \\
&&+ \Lambda_1 (\psi^{A\dagger}_+ v_t^+ \psi^A_- + 
\psi^{A\dagger}_- v_t^- \psi^A_+ + \psi^{B\dagger}_+ v_t^+ \psi^B_- + 
\psi^{B\dagger}_- v_t^- \psi^B_+) \nonumber \\
&&+ \Lambda_2 (\psi^{A\dagger}_+ v_t^+ \psi^B_- + 
\psi^{B\dagger}_- v_t^- \psi^A_+ - \psi^{B\dagger}_+ v_t^+ \psi^A_- - 
\psi^{A\dagger}_- v_t^- \psi^B_+).
\end{eqnarray}
Here $V_t$ is the field defined in eq.(\ref{Vfield}) and the covariant 
derivatives are those of eq.(\ref{covarder}). In components they take the form
\begin{eqnarray}
&&D_\mu \psi^{A,B}_\pm(x) = (\p_\mu \pm i v_\mu^3(x)) \psi^{A,B}_\pm(x),
\nonumber \\
&&D_\mu \psi^{A,B\dagger}_\pm(x) = 
(\p_\mu \mp i v_\mu^3(x)) \psi^{A,B\dagger}_\pm(x).
\end{eqnarray}
Note that $v_t^3$ as well as $v_t^\pm$ (and hence $V_t$) count like one 
temporal derivative because these composite fields indeed contain one 
time-derivative of the magnon field. 

When one derives the Euclidean path integral from the Hamiltonian formulation 
of the effective theory, the term
$\psi^{A\dagger}_+ \p_t \psi^A_+ + \psi^{A\dagger}_- \p_t \psi^A_- + 
\psi^{B\dagger}_+ \p_t \psi^B_+ + \psi^{B\dagger}_- \p_t \psi^B_-$
arises from the pairs of anticommuting fermion operators. It should be noted 
that there are two more $SU(2)_Q$-breaking but $U(1)_Q$-invariant terms with a 
single time-derivative
\begin{eqnarray}
&&\frac{1}{2} \mbox{Tr}[\Psi^{A\dagger} \sigma_3 D_t \Psi^A \sigma_3 - 
\Psi^{B\dagger} \sigma_3 D_t \Psi^B \sigma_3] \nonumber \\
&&\hspace{2cm}= \psi^{A\dagger}_+ D_t \psi^A_+ - \psi^{A\dagger}_- D_t \psi^A_-
- \psi^{B\dagger}_+ D_t \psi^B_+ + \psi^{B\dagger}_- D_t \psi^B_-, \nonumber \\
&&\frac{1}{2} \mbox{Tr}[\Psi^{A\dagger} D_t \Psi^B \sigma_3 + 
\Psi^{B\dagger} D_t \Psi^A \sigma_3] \nonumber \\
&&\hspace{2cm}= \psi^{A\dagger}_+ D_t \psi^B_+ + \psi^{A\dagger}_- D_t \psi^B_-
+ \psi^{B\dagger}_+ D_t \psi^A_+ + \psi^{B\dagger}_- D_t \psi^A_-.
\end{eqnarray}
These terms need not be included in the effective Lagrangian, since they would 
not imply canonical anticommutation relations in the Hamiltonian formulation. 
In any case, as discussed in appendix B, if one does include these terms they 
can again be removed by an appropriate field redefinition. 

Interestingly, there is only one more term that violates the 
$SU(2)_Q$ symmetry but still respects the $U(1)_Q$ symmetry
\begin{eqnarray}
\widetilde{\cal L}_{1,0,2}
&=&\lambda \mbox{Tr}[\Psi^{A\dagger} V_t \Psi^B \sigma_3] \nonumber \\
&=&\lambda (\psi^{A\dagger}_+ v_t^+ \psi^B_- + 
\psi^{B\dagger}_- v_t^- \psi^A_+ + \psi^{B\dagger}_+ v_t^+ \psi^A_- + 
\psi^{A\dagger}_- v_t^- \psi^B_+).
\end{eqnarray}
Further potential contributions are absent because, for example,
\begin{eqnarray}
&&\mbox{Tr}[\Psi^{A\dagger} D_t \Psi^A \sigma_3] =
\mbox{Tr}[\Psi^{B\dagger} D_t \Psi^B \sigma_3] = 0, \nonumber \\
&&\mbox{Tr}[\Psi^{A\dagger} V_t \Psi^A \sigma_3] =  
\mbox{Tr}[\Psi^{B\dagger} V_t \Psi^B \sigma_3] = 0, \nonumber \\
&&\mbox{Tr}[\Psi^{A\dagger} \sigma_3 V_t \Psi^A \sigma_3] =
\mbox{Tr}[\Psi^{B\dagger} \sigma_3 V_t \Psi^B \sigma_3] = 0.
\end{eqnarray}

Terms with a single spatial derivative are forbidden due to the reflection 
symmetry $R$ and the 90 degrees rotation symmetry $O$ of the quadratic spatial 
lattice of the underlying microscopic system. The terms with two spatial 
derivatives are given by
\begin{eqnarray}
{\cal L}_{0,2,2}&=&\frac{1}{2 M_1'} \mbox{Tr}[D_i \Psi^{A\dagger} D_i \Psi^B] 
+ \frac{1}{4 M_2'} \mbox{Tr}[D_i \Psi^{A\dagger} \sigma_3 D_i \Psi^A - 
D_i \Psi^{B\dagger} \sigma_3 D_i \Psi^B] \nonumber \\
&&+ i K_1 \mbox{Tr}[D_i \Psi^{A\dagger} V_i \Psi^B + 
D_i \Psi^{B\dagger} V_i \Psi^A] \nonumber \\
&&+ i K_2 \mbox{Tr}[D_i \Psi^{A\dagger} \sigma_3 V_i \Psi^A -
D_i \Psi^{B\dagger} \sigma_3 V_i \Psi^B] \nonumber \\
&&+ N_1 \mbox{Tr}[\Psi^{A\dagger} V_i V_i \Psi^B]
+ \frac{N_2}{2} \mbox{Tr}[\Psi^{A\dagger} V_i \sigma_3 V_i \Psi^A - 
\Psi^{B\dagger} V_i \sigma_3 V_i \Psi^B] \nonumber \\
&=& \frac{1}{2M_1'}(D_i\pAPD D_i\pBP + D_i\pBPD D_i\pAP + D_i\pAMD D_i \pBM + 
D_i\pBMD D_i\pAM) \nonumber \\
&&+  \frac{1}{2M_2'}(D_i\pAPD D_i\pAP - D_i\pAMD D_i\pAM - D_i\pBPD D_i\pBP + 
D_i\pBMD D_i\pBM) \nonumber \\
&&+ i K_1(D_i\pAPD v_i^+ \pBM - \pBMD v_i^- D_i\pAP + D_i\pAMD v_i^-\pBP - 
\pBPD v_i^+ D_i\pAM \nonumber \\
&&\qquad + D_i\pBPD v_i^+ \pAM - \pAMD v_i^- D_i\pBP + D_i\pBMD v_i^-\pAP - 
\pAPD v_i^+ D_i\pBM) \nonumber \\
&&+ i K_2(D_i\pAPD v_i^+\pAM - \pAMD v_i^- D_i\pAP - D_i\pAMD v_i^-\pAP + 
\pAPD v_i^+ D_i\pAM \nonumber \\
&&\qquad - D_i\pBPD v_i^+\pBM + \pBMD v_i^- D_i\pBP + D_i\pBMD v_i^-\pBP - 
\pBPD v_i^+ D_i\pBM) \nonumber \\
&&+ N_1(\pAPD v_i^+v_i^-\pBP + \pAMD v_i^-v_i^+\pBM + 
\pBPD v_i^+v_i^-\pAP + \pBMD v_i^-v_i^+\pAM)  \nonumber \\
&&- N_2(\pAPD v_i^+v_i^-\pAP - \pAMD v_i^-v_i^+\pAM -
 \pBPD v_i^+v_i^-\pBP + \pBMD v_i^-v_i^+\pBM).
\end{eqnarray}
Note that the imaginary unit $i$ in front of the terms proportional to $K_1$ 
and $K_2$ is necessary to ensure that the corresponding Hamiltonian is 
Hermitean. In principle, terms containing $D_i D_i$ and $D_i V_i$ could also be
written down. However, upon partial integration, up to irrelevant surface terms
they lead to the same Euclidean action as the terms constructed here. Since the
doped electrons or holes are non-relativistic, there is no reason why the 
kinetic mass parameters $M_1'$ and $M_2'$ should agree with the rest mass 
parameters $M_1$ and $M_2$. In addition, there are again terms that break the 
$SU(2)_Q$ symmetry but leave the $U(1)_Q$ symmetry intact
\begin{eqnarray}
\widetilde{\cal L}_{0,2,2}
&=&\frac{1}{4 m'} \mbox{Tr}[D_i \Psi^{A\dagger} D_i \Psi^A \sigma_3 + 
D_i \Psi^{B\dagger} D_i \Psi^B \sigma_3] \nonumber \\
&&+ i \kappa_1 \mbox{Tr}[D_i \Psi^{A\dagger} \sigma_3 V_i \Psi^B 
\sigma_3 + D_i \Psi^{B\dagger} V_i \sigma_3 \Psi^A \sigma_3] \nonumber \\
&&+ i \kappa_2 \mbox{Tr}[D_i \Psi^{A\dagger} V_i \Psi^A \sigma_3 + 
D_i \Psi^{B\dagger} V_i \Psi^B \sigma_3] \nonumber \\
&&+ \frac{\nu}{2} \mbox{Tr}[\Psi^{A\dagger} V_i V_i \Psi^A \sigma_3 + 
\Psi^{B\dagger} V_i V_i \Psi^B \sigma_3] \nonumber \\
&=& \frac{1}{2m'}(D_i\pAPD D_i\pAP + D_i\pAMD D_i\pAM + D_i\pBPD D_i\pBP + 
D_i\pBMD D_i\pBM) \nonumber \\
&&+ i\kappa_1 (D_i\pAPD v_i^+\pBM - \pBMD v_i^- D_i\pAP - D_i\pAMD v_i^-\pBP + 
\pBPD v_i^+ D_i\pAM  \nonumber \\
&&\qquad - D_i\pBPD v_i^+\pAM + \pAMD v_i^- D_i\pBP + D_i\pBMD v_i^-\pAP - 
\pAPD v_i^+ D_i\pBM) \nonumber \\
&&+ i\kappa_2 (D_i\pAPD v_i^+\pAM - \pAMD v_i^- D_i\pAP + D_i\pAMD v_i^-\pAP -
\pAPD  v_i^+ D_i\pAM \nonumber \\
&&\qquad + D_i\pBPD v_i^+\pBM - \pBMD v_i^- D_i\pBP + D_i\pBMD v_i^-\pBP -
\pBPD v_i^+ D_i\pBM) \nonumber \\
&&+ \nu(\pAPD v_i^+v_i^-\pAP + \pAMD v_i^-v_i^+\pAM + \pBPD v_i^+v_i^-\pBP + 
\pBMD v_i^-v_i^+\pBM).
\end{eqnarray}

Next we consider terms quartic in the fermion fields which describe short-range
interactions between the charge carriers. To lowest order there are five
linearly independent 4-fermion contact interaction terms
\begin{eqnarray}
{\cal L}_{0,0,4}
&=&\frac{G_1}{12} \mbox{Tr}[\Psi^{A\dagger} \Psi^A \Psi^{A\dagger} \Psi^A +
\Psi^{B\dagger} \Psi^B \Psi^{B\dagger} \Psi^B] + 
\frac{G_2}{2} \mbox{Tr}[\Psi^{A\dagger} \Psi^A \Psi^{B\dagger} \Psi^B] 
\nonumber \\
&&+ \frac{G_3}{2} \mbox{Tr}[\Psi^{A\dagger} \Psi^B \Psi^{B\dagger} \Psi^A]
+ \frac{G_4}{2} \mbox{Tr}[\Psi^{A\dagger} \sigma_3 \Psi^A \Psi^{B\dagger} 
\sigma_3 \Psi^B] \nonumber \\
&&+ G_5 \mbox{Tr}[\Psi^{A\dagger} \sigma_3 \Psi^A \Psi^{A\dagger} \Psi^B - 
\Psi^{B\dagger} \sigma_3 \Psi^B \Psi^{B\dagger} \Psi^A] \nonumber \\
&=&G_1(\pAPD\pAP\pAMD\pAM + \pBPD\pBP\pBMD\pBM) \nonumber \\
&&+ G_2(\pAPD\pAP\pBPD\pBP + \pAPD\pAP\pBMD\pBM + \pAMD\pAM\pBPD\pBP +
\pAMD\pAM\pBMD\pBM \nonumber \\
&&\qquad - 2 \pAPD\pBP\pAMD\pBM - 2 \pBPD\pAP\pBMD\pAM) \nonumber \\
&&+ G_3(\pAPD\pAP\pBMD\pBM + \pAMD\pAM\pBPD\pBP - \pAPD\pAP\pBPD\pBP 
- \pAMD\pAM\pBMD\pBM \nonumber \\
&&\qquad - 2 \pAPD\pAM\pBMD\pBP - 2 \pAMD\pAP\pBPD\pBM) \nonumber \\
&&+ G_4(\pAPD\pAP\pBPD\pBP - \pAPD\pAP\pBMD\pBM - \pAMD\pAM\pBPD\pBP +
\pAMD\pAM\pBMD\pBM) \nonumber \\
&&+ G_5(\pAPD\pAP\pAMD\pBM + \pAPD\pAP\pBMD\pAM - \pBPD\pBP\pAMD\pBM -
\pBPD\pBP\pBMD\pAM \nonumber \\
&&\qquad - \pAPD\pBP\pAMD\pAM - \pBPD\pAP\pAMD\pAM + \pAPD\pBP\pBMD\pBM +
\pBPD\pAP\pBMD\pBM). \nonumber \\
\end{eqnarray}
It is interesting to note that
\begin{eqnarray}
&&\mbox{Tr}[\Psi^{A\dagger} \Psi^B \Psi^{B\dagger} \Psi^A +
\Psi^{A\dagger} \Psi^A \Psi^{B\dagger} \Psi^B + 
2 \Psi^{A\dagger} \Psi^B \Psi^{A\dagger} \Psi^B] = 0,
\nonumber \\
&&\mbox{Tr}[\Psi^{A\dagger} \Psi^B \Psi^{B\dagger} \Psi^A +
\Psi^{A\dagger} \sigma_3 \Psi^B \Psi^{B\dagger} \sigma_3 \Psi^A + 
2 \Psi^{A\dagger} \sigma_3 \Psi^A \Psi^{B\dagger} \sigma_3 \Psi^B] = 0,
\nonumber \\
&&\mbox{Tr}[\Psi^{A\dagger} \Psi^B \Psi^{B\dagger} \Psi^A -
\Psi^{A\dagger} \Psi^A \Psi^{B\dagger} \Psi^B] =
2 (\mbox{Tr}[\Psi^{A\dagger} \Psi^B])^2.
\end{eqnarray}
Together with other relations similar to these ones, this implies that the 
terms listed above form a maximal linearly independent set. 

Again, there are additional terms that are invariant under $U(1)_Q$ but not 
under $SU(2)_Q$
\begin{eqnarray}
\widetilde{\cal L}_{0,0,4}\!\!\!&=&\!\!\!
\frac{g_1}{4} 
\mbox{Tr}[\Psi^{A\dagger} \sigma_3 \Psi^A \Psi^{B\dagger} \Psi^B \sigma_3
- \Psi^{B\dagger} \sigma_3 \Psi^B \Psi^{A\dagger} \Psi^A \sigma_3] +
\frac{g_2}{2} 
\mbox{Tr}[\Psi^{A\dagger} \Psi^A \sigma_3 \Psi^{B\dagger} \Psi^B \sigma_3]
\nonumber \\
&&+ g_3 \mbox{Tr}[\Psi^{A\dagger} \Psi^A \sigma_3 \Psi^{A\dagger} \Psi^B + 
\Psi^{B\dagger} \Psi^B \sigma_3 \Psi^{B\dagger} \Psi^A] \nonumber \\
&=&g_1(\pAPD\pAP\pBMD\pBM - \pAMD\pAM\pBPD\pBP) \nonumber \\
&&+ g_2(\pAPD\pAP\pBPD\pBP + \pAPD\pAP\pBMD\pBM + \pAMD\pAM\pBPD\pBP +
\pAMD\pAM\pBMD\pBM \nonumber \\
&&\qquad + 2 \pAPD\pBP\pAMD\pBM + 2 \pBPD\pAP\pBMD\pAM)\nonumber \\
&&- g_3(\pAPD\pBP\pAMD\pAM + \pAPD\pBP\pBMD\pBM + \pAMD\pBM\pAPD\pAP +
\pAMD\pBM\pBPD\pBP \nonumber \\
&&\qquad + \pBPD\pAP\pAMD\pAM + \pBPD\pAP\pBMD\pBM + \pBMD\pAM\pAPD\pAP +
\pBMD\pAM\pBPD\pBP). \nonumber \\
\end{eqnarray}
One may note that, for example,
\begin{equation}
\mbox{Tr}[\Psi^{A\dagger} \Psi^A \Psi^{B\dagger} \Psi^B +
\Psi^{A\dagger} \Psi^A \sigma_3 \Psi^{B\dagger} \Psi^B \sigma_3 + 
2 \Psi^{A\dagger} \Psi^B \sigma_3 \Psi^{B\dagger} \Psi^A \sigma_3] = 0.
\end{equation}
Together with further relations of a similar kind, this implies that there are 
no other linearly independent 4-fermion terms that obey the relevant 
symmetries.

For completeness, let us also construct the terms containing six fermion fields
and no derivatives. The $SU(2)_Q$-invariant 6-fermion terms can be written as
\begin{eqnarray}
{\cal L}_{0,0,6}
&=&\frac{H_1}{4} \mbox{Tr}[\Psi^{A\dagger} \sigma_3 \Psi^A 
\Psi^{A\dagger} \sigma_3 \Psi^A \Psi^{B\dagger} \sigma_3 \Psi^B -
\Psi^{B\dagger} \sigma_3 \Psi^B 
\Psi^{B\dagger} \sigma_3 \Psi^B \Psi^{A\dagger} \sigma_3 \Psi^A] \nonumber \\
&&+ \frac{H_2}{3} \mbox{Tr}[\Psi^{A\dagger} \Psi^B \Psi^{A\dagger} \Psi^B
\Psi^{A\dagger} \Psi^B] \nonumber \\
&=&H_1(\pAPD\pAP\pBPD\pBP\pBMD\pBM + \pAPD\pAP\pAMD\pAM\pBMD\pBM \nonumber \\
&&\qquad - \pAMD\pAM\pBPD\pBP\pBMD\pBM - \pAPD\pAP\pAMD\pAM\pBPD\pBP) 
\nonumber \\
&&+ H_2(\pAPD\pAP\pBPD\pBP\pBMD\pAM + \pAMD\pAM\pBMD\pBM\pBPD\pAP \nonumber \\
&&\qquad + \pAMD\pAM\pBMD\pBM\pAPD\pBP + \pAPD\pAP\pBPD\pBP\pAMD\pBM).
\end{eqnarray}
It is interesting to note that
\begin{eqnarray}
&&\mbox{Tr}[\Psi^{A\dagger} \sigma_3 \Psi^A 
\Psi^{A\dagger} \sigma_3 \Psi^A \Psi^{B\dagger} \sigma_3 \Psi^B -
\Psi^{B\dagger} \sigma_3 \Psi^B 
\Psi^{B\dagger} \sigma_3 \Psi^B \Psi^{A\dagger} \sigma_3 \Psi^A] \nonumber \\
&&\hspace{1.5cm}= \frac{1}{2} 
\mbox{Tr}[\Psi^{A\dagger} \sigma_3 \Psi^A - \Psi^{B\dagger} \sigma_3 \Psi^B]
\mbox{Tr}[\Psi^{A\dagger} \sigma_3 \Psi^A \Psi^{B\dagger} \sigma_3 \Psi^B].
\end{eqnarray}
In addition, there is one $SU(2)_Q$-breaking (but $U(1)_Q$-invariant) 6-fermion
term
\begin{eqnarray}
\widetilde{\cal L}_{0,0,6}\!\!\!&=&\!\!\!\frac{h}{4} 
\mbox{Tr}[\Psi^{A\dagger} \Psi^A \sigma_3 \Psi^{A\dagger} \Psi^A \sigma_3 
\Psi^{B\dagger} \Psi^B \sigma_3 + \Psi^{B\dagger} \Psi^B \sigma_3 
\Psi^{B\dagger} \Psi^B \sigma_3 \Psi^{A\dagger} \Psi^A \sigma_3] \nonumber \\
&=&- h(\pAPD\pAP\pAMD\pAM\pBPD\pBP + \pAPD\pAP\pAMD\pAM\pBMD\pBM \nonumber \\
&&\qquad + \pAPD\pAP\pBPD\pBP\pBMD\pBM + \pAMD\pAM\pBPD\pBP\pBMD\pBM).
\end{eqnarray}
Finally, the only 8-fermion term with no derivatives takes the form
\begin{eqnarray}
{\cal L}_{0,0,8}
&=&\frac{I}{24} \ \mbox{Tr}[\Psi^{A\dagger} \Psi^A \Psi^{B\dagger} \Psi^B 
\Psi^{A\dagger} \Psi^A \Psi^{B\dagger} \Psi^B] \nonumber \\ 
&=&- I \ \pAPD\pAP\pBPD\pBP\pAMD\pAM\pBMD\pBM,
\end{eqnarray}
which is $SU(2)_Q$-invariant. It may be noted that
\begin{equation}
\mbox{Tr}[\Psi^{A\dagger} \Psi^A \Psi^{B\dagger} \Psi^B 
\Psi^{A\dagger} \Psi^A \Psi^{B\dagger} \Psi^B] + \frac{1}{2}
(\mbox{Tr}[\Psi^{A\dagger} \Psi^A \Psi^{B\dagger} \Psi^B])^2 = 0.
\end{equation}
No $SU(2)_Q$-breaking 8-fermion term without derivatives exists, such that 
$\widetilde{\cal L}_{0,0,8} = 0$. Terms with more than eight fermion fields 
vanish due to the Pauli principle, unless one includes derivatives. Since such 
terms are of higher order than those without derivatives, they will not be 
constructed here. When one wants to address questions for which the 
short-distance forces between charge carriers are essential, it will be 
necessary to consider such terms. While constructing them is a straightforward
exercise, it is not very illuminating and will hence be omitted at this stage.

The 4-, 6-, and 8-fermion contact terms parameterize short distance 
interactions with a large number of undetermined low-energy constants. Since
this limits the predictive power of the effective theory at short distances, it
is natural to concentrate on long-distance forces between the charge carriers.
For example, one-magnon exchange mediates a long-range force that is 
unambiguously predicted by the effective theory in terms of just a few 
low-energy parameters.

It should be mentioned that there are many equivalent ways of rewriting the
various contributions to the action in terms of traces. Hence, the above 
choices of terms are to some extent arbitrary. It is important that the 
selected terms form a maximal linearly independent set. For example, all 
determinants or products of traces of fermion fields can be written as linear
combinations of the traces listed above. To verify the completeness and 
linear independence of the selected terms is a non-trivial task which was 
addressed by extensive use of the algebraic manipulation program FORM 
\cite{Ver00}.

It is straightforward to include the fermion chemical potential $\mu$ in the 
effective theory. It appears as the temporal component of a purely imaginary 
$U(1)_Q$ gauge field and thus manifests itself in an additional contribution to
the covariant derivative
\begin{equation}
D_t \Psi^{A,B}(x) = \p_t \Psi^{A,B}(x) + i v_t^3(x) \sigma_3 \Psi^{A,B}(x) -
\mu \Psi^{A,B}(x) \sigma_3.
\end{equation}
Before one can do consistent loop-calculations in the low-energy effective 
theory at non-zero $Q$ or at non-zero $\mu$ one must develop a power-counting 
scheme, e.g.\ along the lines of \cite{Bec99}. This will be the subject of a 
future publication.

\subsection{Dispersion Relations of Electrons and Holes}

In this subsection, as an application of the effective theory, we consider the 
dispersion relations of the charge carriers. For this purpose we switch off the
magnon field (i.e.\ $P(x) = \frac{1}{2}(\1 + \sigma_3) \ \Rightarrow \
u(x) = \1, v_\mu(x) = 0$) and consider the propagation of free charge carriers 
in the antiferromagnetic medium. In the absence of $SU(2)_Q$-breaking terms, 
the Lagrangian (quadratic in the fermion fields) then reduces to
\begin{eqnarray}
{\cal L}&=&\psi^{A\dagger}_+ \p_t \psi^A_+ + \psi^{A\dagger}_- \p_t \psi^A_- +
\psi^{B\dagger}_+ \p_t \psi^B_+ + \psi^{B\dagger}_- \p_t \psi^B_- \nonumber \\
&&+ (\psi^{A\dagger}_+, \psi^{B\dagger}_+) \left(\begin{array}{cc}
M_2 & M_1 \\ M_1 & - M_2 \end{array}\right)
\left(\begin{array}{c} \psi^A_+ \\ \psi^B_+ \end{array}\right) 
+ (\psi^{A\dagger}_-, \psi^{B\dagger}_-) \left(\begin{array}{cc}
- M_2 & M_1 \\ M_1 & M_2 \end{array}\right)
\left(\begin{array}{c} \psi^A_- \\ \psi^B_- \end{array}\right) \nonumber \\
&&+ (\p_i \psi^{A\dagger}_+, \p_i \psi^{B\dagger}_+) \left(\begin{array}{cc}
\frac{1}{2 M'_2} & \frac{1}{2 M'_1} \\ 
\frac{1}{2 M'_1} & - \frac{1}{2 M'_2} \end{array}\right)
\left(\begin{array}{c} \p_i \psi^A_+ \\ \p_i \psi^B_+ \end{array}\right) 
\nonumber \\
&&+ (\p_i \psi^{A\dagger}_-, \p_i \psi^{B\dagger}_-) \left(\begin{array}{cc}
- \frac{1}{2 M'_2} & \frac{1}{2 M'_1} \\ 
\frac{1}{2 M'_1} & \frac{1}{2 M'_2} \end{array}\right)
\left(\begin{array}{c} \p_i \psi^A_- \\ \p_i \psi^B_- \end{array}\right).
\end{eqnarray}
The eigenstates of free particles propagating with a 2-d momentum vector 
$\vec p$ arise as the eigenvectors of the matrices
\begin{eqnarray}
H_+(p^2)&=&\left(\begin{array}{cc} M_2 + \frac{p^2}{2 M'_2} & 
M_1 + \frac{p^2}{2 M'_1} \\ M_1 + \frac{p^2}{2 M'_1} & 
- M_2 - \frac{p^2}{2 M'_2} \end{array}\right), \nonumber \\
H_-(p^2)&=&\left(\begin{array}{cc} - M_2 - \frac{p^2}{2 M'_2} & 
M_1 + \frac{p^2}{2 M'_1} \\ M_1 + \frac{p^2}{2 M'_1} & 
M_2 + \frac{p^2}{2 M'_2} \end{array}\right).
\end{eqnarray}
Due to the lack of Galilean invariance the eigenvectors depend on $p^2$, i.e.\
the probability for an electron or hole to be found on the $A$ or $B$ 
sublattice depends on the momentum. As a consequence of the displacement 
symmetries $D$ and $D'$ the eigenvalues of $H_+(p^2)$ and $H_-(p^2)$ are the 
same. Both matrices have two eigenvalues
\begin{eqnarray}
E_{1,2}(p^2)&=&\pm \sqrt{\left(M_1 + \frac{p^2}{2 M'_1}\right)^2 +
\left(M_2 + \frac{p^2}{2 M'_2}\right)^2} \nonumber \\
&=&\pm \left(M + \frac{p^2}{2 M'} + {\cal O}(p^4)\right).
\end{eqnarray}
The positive energy states correspond to electrons, while the negative energy
states correspond to holes. Not surprisingly, due to the $SU(2)_Q$ symmetry
electrons and holes have the same dispersion relation. The rest mass $M$ and
the kinetic mass $M'$ are given by
\begin{equation}
M = \sqrt{M_1^2 + M_2^2}, \quad
\frac{M}{M'} = \frac{M_1}{M'_1} + \frac{M_2}{M'_2}.
\end{equation}

Next we take into account the additional terms that reduce the $SU(2)_Q$ 
symmetry to the $U(1)_Q$ symmetry. Then there are additional contributions to 
the energy
\begin{equation}
\widetilde H_+(p^2) = \widetilde H_-(p^2) =
\left(\begin{array}{cc} m + \frac{p^2}{2 m'} & 0 \\ 0 & 
m + \frac{p^2}{2 m'} \end{array}\right).
\end{equation}
and the corresponding eigenvalues now take the form
\begin{equation}
E_{1,2}(p^2) = m + \frac{p^2}{2 m'} \pm 
\left(M + \frac{p^2}{2 M'}\right) + {\cal O}(p^4).
\end{equation}
Still, the energies in the $+$ and $-$ sectors are the same. However, the
electron and hole dispersion relations now differ. 

At this point, we have constructed eigenstates of the free Hamiltonian with 
definite continuum momentum and with definite spin projection on the direction 
of the staggered magnetization. However, unlike the eigenstates of the 
underlying microscopic Hamiltonian, the states of the effective theory do not 
have a definite lattice momentum. Still, the low-energy effective theory 
defined in the continuum knows about the underlying lattice structure through
the realization of the displacement symmetries $D$ and $D'$. Since the symmetry
$D$ is spontaneously broken, neither the vacuum nor the single particle states
are eigenstates of $D$. Operating twice with $D$ acts trivially on the fields,
i.e.\ $^{DD} P(x) = P(x)$, $^{DD} \Psi^{A,B}_\pm = \Psi^{A,B}_\pm$, and hence
does not reveal any useful information. It is more useful to operate with the 
unbroken displacement symmetry $D'$. In particular, the vacuum state 
$P(x) = \frac{1}{2}(\1 + \sigma_3)$ is invariant under $D'$. Still, in the way 
we constructed them, the electron or hole states of the effective theory are 
not eigenstates of $D'$. However, since states with spin parallel and
antiparallel to the staggered magnetization are degenerate with each other,
one can form appropriate linear combinations that are eigenstates of the 
displacement symmetry $D'$. Applying $D'$ twice one obtains
\begin{equation}
^{D'D'}\psi^{A,B}_\pm(x) = \pm \ ^{D'}\psi^{B,A}_\mp(x) = - \psi^{A,B}_\pm(x),
\end{equation}
which implies that the corresponding eigenvalue $\lambda = \exp(i k a)$ of $D'$
obeys 
\begin{equation}
\lambda^2 = \exp(2 i k a) = - 1 \ \Rightarrow \ k a = \pm \frac{\pi}{2}.
\end{equation}
This is reminiscent of the result, mentioned in the introduction, that 
low-energy hole states are located at lattice momenta 
$(\pm \frac{\pi}{2},\pm \frac{\pi}{2})$
\cite{Wel95,LaR97,Kim98,Ron98,Tru88,Shr88,Els90,Bru00}. However, the comparison
with these findings is subtle. In particular, the results of the exact 
diagonalization study on small \cite{Els90} and of the Monte Carlo study on
larger volumes \cite{Bru00} must be interpreted carefully. In a finite volume
(with periodic boundary conditions), in analogy to QCD \cite{Leu87}, both the
$SU(2)_s$ spin symmetry and the displacement symmetry $D$ are restored and the 
staggered magnetization acts as a quantum rotor \cite{Has93}. As a result, in 
contrast to the infinite volume limit, the single particle states in a finite 
volume can be constructed as eigenstates of $D$. It is interesting to note that
the finite volume effects that lead to the restoration of the spontaneously 
broken symmetries $SU(2)_s$ and $D$ can be understood in the framework of the 
effective theory. This requires a nonperturbative quantum mechanical treatment 
along the lines of \cite{Leu87,Has93}.

\section{Systems with Holes only}

In this section we consider the $t$-$J$ model as well as its low-energy 
effective theory. In the $t$-$J$ model holes are the only charge carriers 
which leads to substantial simplifications in the effective theory.

\subsection{The $t$-$J$ Model}

The $t$-$J$ model is defined by the Hamilton operator
\begin{equation}
H = P \left\{- t \sum_{x,i} 
(c_x^\dagger c_{x+\hat i} + c_{x+\hat i}^\dagger c_x) +
J \sum_{x,i} \vec S_x \cdot \vec S_{x+\hat i} - \mu \sum_x (n_x - 1)\right\} P,
\end{equation}
with
\begin{equation}
c_x = \left(\begin{array}{c} c_{x\uparrow} \\ c_{x\downarrow} \end{array}
\right), \quad
S_x = c_x^\dagger \frac{\vec \sigma}{2} c_x, \quad n_x = c_x^\dagger c_x.
\end{equation}
In contrast to the Hubbard model, in the $t$-$J$ model the operators act in a
restricted Hilbert space of empty or at most singly occupied sites. In 
particular, states with doubly occupied sites are exiled from the physical
Hilbert space by the projection operator $P$. Hence, by definition, the $t$-$J$
model does not allow the addition of electrons to a half-filled state. 
Consequently, the only charge carriers are holes.

It is straightforward to show that the $t$-$J$ model has the same symmetries as
the Hubbard model. The only exception is the $SU(2)_Q$ symmetry which relates
electrons to holes in the Hubbard model, and which is absent in the $t$-$J$ 
model. Still, the Abelian fermion number symmetry $U(1)_Q$ remains exact in the
$t$-$J$ model.

\subsection{Effective Theory for Magnons and Holes}

Since, up to the $SU(2)_Q$ symmetry, the $t$-$J$ model has the same symmetries 
as the Hubbard model, the effective theory of the previous section also applies
in this case. Of course, the values of the low-energy parameters will be 
different than for the Hubbard model. Still, the absence of electrons beyond 
half-filling leads to drastic simplifications. In particular, in the effective 
theory the absence of electrons manifests itself by an infinite electron rest 
mass. Consequently, with a finite amount of energy these excitations cannot be 
generated. As discussed in the previous section, the diagonalization of the 
mass matrices of electrons and holes yields
\begin{eqnarray}
&&U_\pm \left(\begin{array}{cc} m \pm M_2 & M_1 \\ M_1 & m \mp M_2 \end{array} 
\right) U_\pm^\dagger = \left(\begin{array}{cc} m \pm \sqrt{M_1^2 + M_2^2} & 0 
\\ 0 &  m \mp \sqrt{M_1^2 + M_2^2} \end{array} \right), \nonumber \\
&&U_\pm = \left(\begin{array}{cc} X & \pm Y \\ \mp Y & X \end{array} \right), 
\quad X, Y \in \R.
\end{eqnarray}
The eigenvectors corresponding to the eigenvalue $m + \sqrt{M_1^2 + M_2^2}$
describe electrons, while the ones corresponding to $m - \sqrt{M_1^2 + M_2^2}$ 
describe holes. When the electron rest mass $m + \sqrt{M_1^2 + M_2^2}$ goes to
infinity, the corresponding eigenvector fields
\begin{equation}
X \psi^A_+(x) + Y \psi^B_+(x) = 0, \quad 
Y \psi^A_-(x) + X \psi^B_-(x) = 0,
\end{equation}
which describe electrons, must be put to zero. The orthogonal combinations
\begin{equation}
\psi_+(x) = - Y \psi^A_+(x) + X \psi^B_+(x), \quad 
\psi_-(x) = X \psi^A_-(x) - Y \psi^B_-(x),
\end{equation}
describe holes and must be kept. As a result, the number of degrees of 
freedom is reduced by a factor of two. In complete analogy to the discussion in
appendix B one can show that the hole field $\psi_\pm(x)$ transforms as follows
under the various symmetry operations
\begin{eqnarray}
SU(2)_s:&&\psi_\pm(x)' = \exp(\pm i \alpha(x)) \psi_\pm(x), \nonumber \\
U(1)_Q:&&^Q\psi_\pm(x) = \exp(i \omega) \psi_\pm(x), \nonumber \\
D:&&^D\psi_\pm(x) = \mp \exp(\mp i \varphi(x)) \psi_\mp(x), \nonumber \\
D':&&^{D'}\psi_\pm(x) = \pm \psi_\mp(x), \nonumber \\
O:&&^O\psi_\pm(x) = \psi_\pm(Ox), \nonumber \\
R:&&^R\psi_\pm(x) = \psi_\pm(Rx), \nonumber \\
T:&&^T\psi_\pm(x) =  \exp(\mp i \varphi(Tx)) \psi^\dagger_\pm(Tx), \nonumber \\
&&^T\psi^\dagger_\pm(x) = - \exp(\pm i \varphi(Tx)) \psi_\pm(Tx), \nonumber \\
T':&&^{T'}\psi_\pm(x) = - \psi^\dagger_\pm(Tx), \nonumber \\
&&^{T'}\psi^\dagger_\pm(x) = \psi_\pm(Tx).
\end{eqnarray}
Hence, except for the $SU(2)_Q$ symmetry, all symmetries can also be 
implemented on the hole fields alone. It should be noted that the 
transformation laws for $\psi_\pm(x)$ result from those for $\psi^{A,B}_\pm(x)$
simply by dropping the sublattice indices $A$ and $B$.

The absence of electron fields also drastically reduces the number of terms one
can write down in the low-energy effective theory. In particular, the leading 
terms in the effective action now take the form
\begin{eqnarray}
S[\psi^\dagger_\pm,\psi_\pm,P]&=&\int d^2x \ dt \
\{\rho_s \mbox{Tr}[\p_i P \p_i P + \frac{1}{c^2} \p_t P \p_t P] 
+ M (\psi_+^\dagger \psi_+ + \psi_-^\dagger \psi_-) \nonumber \\
&&+ \psi_+^\dagger D_t \psi_+ + \psi_-^\dagger D_t \psi_- +
\frac{1}{2 M'} (D_i\psi^\dagger_+ D_i\psi_+ + D_i\psi^\dagger_- D_i\psi_-)
\nonumber \\
&&+ \Lambda (\psi_+^\dagger v^+_t \psi_- + \psi_-^\dagger v^-_t \psi_+) 
\nonumber \\
&&+i K (D_i\psi^\dagger_+ v_i^+ \psi_- - \psi_-^\dagger v_i^- D_i \psi_+ +
D_i\psi^\dagger_- v_i^- \psi_+ - \psi_+^\dagger v_i^+ D_i \psi_-) \nonumber \\
&&+N (\psi_+^\dagger v_i^+ v_i^- \psi_+ + \psi_-^\dagger v_i^- v_i^+ \psi_-) +
G \psi_+^\dagger \psi_+ \psi_-^\dagger \psi_-\}. 
\end{eqnarray}
This form of the effective action is similar to (but not identical with) the 
ones of \cite{Wen89,Sha90,Kue93,Kar98,Liu03}. In particular, in some of those 
works spin-charge separation was invoked and spinless fermions were considered.
Also the role of the sublattice indices (which have at this stage disappeared 
from our description) is different in those approaches. Furthermore, the 
dynamical role attributed to the composite gauge field in some of those works 
is different than in our effective theory. It should be pointed out that the 
above effective Lagrangian correctly describes the low-energy dynamics of holes
only if electrons are completely absent beyond half-filling (as it is indeed 
the case in the $t$-$J$ model). Otherwise the general effective theory of the 
previous section with a larger number of low-energy constants (and thus with 
somewhat reduced predictive power) must be employed.

\section{Coupling to External Electromagnetic Fields}

In the following sections we will couple both microscopic and effective 
theories for antiferromagnets to external electromagnetic fields. For this
purpose, we will make use of an observation by Fr\"ohlich and Studer concerning
the Pauli equation \cite{Fro92}.

\subsection{Local $SU(2)_s$ Symmetry of the Pauli Equation}

Up to corrections of order $1/M_e^3$ (where $M_e$ is the electron mass) the 
Pauli equation (i.e.\ the non-relativistic reduction of the Dirac equation to 
its upper components) takes the form 
\begin{equation}
\label{Pauli}
i (\p_t - i e \Phi + i \frac{e}{8 M_e^2} \vec \nabla \cdot \vec E
+ i \frac{e}{2M_e} \vec B \cdot \vec \sigma) \Psi = 
- \frac{1}{2 M_e}(\vec \nabla + i e \vec A - 
i \frac{e}{4 M_e} \vec E \times \vec \sigma)^2 \Psi.
\end{equation}
Here $\Psi(x)$ is a 2-component Pauli spinor at the space-time point 
$x = (\vec x,t)$, $\vec \sigma$ are the Pauli matrices, $\Phi(x)$ and 
$\vec A(x)$ are the electromagnetic scalar and vector potentials, and 
\begin{equation}
\label{Maxwell}
\vec E(x) = - \vec \nabla \Phi(x) - \p_t \vec A(x), \quad 
\vec B(x) = \vec \nabla \times \vec A(x),
\end{equation}
are the usual electromagnetic field strengths. The first two terms on the 
left-hand side of eq.(\ref{Pauli}) form the $U(1)_Q$ covariant derivative
familiar from QED. The third (Darwin) and fourth (Zeeman) term on the left-hand
side represent relativistic corrections. The first two terms on the right-hand 
side again form an ordinary $U(1)_Q$ covariant derivative, while the third term
represents the relativistic spin-orbit coupling. The Pauli equation transforms 
covariantly under $U(1)_Q$ gauge transformations
\begin{equation}
^Q\Psi(x) = \exp(i \omega(x)) \Psi(x), \quad
^Q\Phi(x) = \Phi(x) + \frac{1}{e} \p_t \omega(x), \quad 
^Q\vec A(x) = \vec A(x) - \frac{1}{e} \vec \nabla \omega(x).
\end{equation}
Obviously, it is also covariant under global spatial rotations
\begin{equation}
^{\cal O}\Psi(\vec x,t) = g \Psi({\cal O} \vec x,t), \quad 
^{\cal O}\Phi(\vec x,t) = \Phi({\cal O} \vec x,t), \quad
^{\cal O}\vec A(\vec x,t) = {\cal O}^T \vec A({\cal O} \vec x,t).
\end{equation}
Here ${\cal O}$ is a general orthogonal $3 \times 3$ rotation matrix with
\begin{equation}
\label{Omatrix}
{\cal O}^T \ \vec \sigma = g^\dagger \vec \sigma g,
\end{equation}
where $g \in SU(2)_s$ represents the rotation ${\cal O} \in SO(3)$ in spinor 
space.

Fr\"ohlich and Studer noticed that the Pauli equation has a hidden local 
$SU(2)_s$ spin symmetry. This symmetry becomes manifest when one writes
\begin{equation}
i D_t \Psi = - \frac{1}{2 M_e} D_i D_i \Psi,
\end{equation}
with the $SU(2)_s \otimes U(1)_Q$ covariant derivative given by
\begin{equation}
\label{covder}
D_\mu = \p_\mu + W_\mu(x) + i e A_\mu(x).
\end{equation}
The components of the non-Abelian vector potential
\begin{equation}
W_\mu(x) = i W_\mu^a(x) \frac{\sigma_a}{2},
\end{equation}
can be identified as the electromagnetic field strengths $\vec E(x)$ and 
$\vec B(x)$, i.e.\
\begin{equation}
\label{Wpot}
W_t^a(x) = \mu_e B^a(x), \quad 
W_i^a(x) = \frac{\mu_e}{2} \varepsilon_{iab} E^b(x).
\end{equation}
The anomalous magnetic moment $\mu_e = g_e e/2 M_e$ of the electron (where, up 
to QED corrections, $g_e = 2$) appears as a non-Abelian gauge coupling. The 
Abelian vector potential $A_\mu(x)$ is the usual one, except for a small 
contribution to the scalar potential due to the Darwin term, 
\begin{equation}
A_t(x) = - \Phi(x) + \frac{1}{8 M_e^2} \vec \nabla \cdot \vec E(x).
\end{equation}
Hence, somewhat unexpected, the Pauli equation also transforms covariantly 
under local $SU(2)_s$ transformations
\begin{equation}
\Psi(x)' = g(x) \Psi(x), \quad 
W_\mu(x)' = g(x) (W_\mu(x) + \p_\mu) g(x)^\dagger.
\end{equation}
It should be pointed out that $SU(2)_s$ is not a gauge symmetry in the usual
sense. In particular, the non-Abelian vector potential $W_\mu(x)$ is not an
independent degree of freedom, but just given in terms of the external 
electromagnetic field strengths $\vec E(x)$ and $\vec B(x)$. The local 
$SU(2)_s$ symmetry is related to the global spatial rotations discussed before.
In particular, global $SU(2)_s$ transformations take the form
\begin{equation}
\Psi(x)' = g \Psi(x), \quad W_\mu(x)' = g W_\mu(x) g^\dagger,
\end{equation}
which, for example, implies
\begin{equation}
\vec B(x)' = {\cal O}^T \vec B(x),
\end{equation}
where the resulting $3 \times 3$ rotation matrix ${\cal O} \in SO(3)$ is again 
given by eq.(\ref{Omatrix}). In contrast to a full spatial rotation, a global 
$SU(2)_s$ transformation does not rotate the argument $\vec x$ of the magnetic
field to ${\cal O} \vec x$. Also the potentials $\Phi(x)$ and $\vec A(x)$ are 
unaffected by the global $SU(2)_s$ symmetry. Consequently, the $SU(2)_s$ 
symmetry is inconsistent with the relations of eq.(\ref{Maxwell}). Despite 
this, the local $SU(2)_s$ symmetry of the Pauli equation, which will be 
inherited by the Hubbard model and by the effective theory, dictates how 
low-frequency external electromagnetic fields are to be included in those 
theories. The high-frequency internal electromagnetic fields (for which 
eq.(\ref{Maxwell}) is essential) are integrated out in the effective theory 
and thus do not spoil the symmetry. The local $SU(2)_s$ structure implies that 
in non-relativistic systems spin plays the role of an internal quantum number 
analogous to flavor in particle physics.

\subsection{The Hubbard Model in an External Electromagnetic Field}

In the next step we want to couple external electromagnetic fields to the
Hubbard model. The Fr\"ohlich-Studer $SU(2)_s$ symmetry of the Pauli equation 
determines how to do this. One must simply use $SU(2)_s \otimes U(1)_Q$ 
covariant derivatives with $\vec E(x)$ and $\vec B(x)$ playing the role of 
non-Abelian vector potentials for $SU(2)_s$. Since the Hubbard model is defined
on a spatial lattice, it is natural to construct corresponding 
$SU(2)_s \otimes U(1)_Q$ parallel transporters ${\cal U}_{x,i}$ connecting 
neighboring lattice sites $x$ and $x + \hat i$,
\begin{equation}
{\cal U}_{x,i} = {\cal P} \exp[\int_0^1 ds \ W_i(x + s \hat i)] \
\exp[i e \int_0^1 ds \ A_i(x + s \hat i)].
\end{equation}
Here ${\cal P}$ denotes path ordering along the link. Under local $SU(2)_s$ 
transformations the parallel transporter transforms as
\begin{equation}
\label{SU2trafo}
{\cal U}_{x,i}' = g(x) \ {\cal U}_{x,i} \ g(x+ \hat i)^\dagger,
\end{equation}
while under $U(1)_Q$ gauge transformations one has
\begin{equation}
\label{U1trafo}
^Q{\cal U}_{x,i} = \exp(i \omega(x)) \ {\cal U}_{x,i} \ 
\exp(- i \omega(x + \hat i)).
\end{equation}
The Hubbard model Hamiltonian coupled to external electromagnetic fields then 
reads
\begin{equation}
H[{\cal U}] = - t \sum_{x, i} (c_x^\dagger {\cal U}_{x,i} c_{x+\hat i} + 
c_{x + \hat i}^\dagger {\cal U}_{x,i}^\dagger c_x)
+ \frac{U}{2} \sum_x (c_x^\dagger c_x - 1)^2 
- \mu \sum_x  (c_x^\dagger c_x - 1),
\end{equation}
and the corresponding Schr\"odinger equation takes the form
\begin{equation}
i D_t \Psi = H[{\cal U}] \Psi.
\end{equation}
Here $\Psi$ is the multi-particle wave function and the covariant derivative is
given by
\begin{equation}
D_t = \p_t + i \sum_x [\vec W_t(x) \cdot \vec S_x + e A_t(x) Q_x].
\end{equation}  
It should be noted that the Zeeman coupling $\mu_e \vec B(x) \cdot \vec S_x$
enters the Hubbard model through $D_t$, while the spin-orbit coupling appears 
in the non-Abelian $SU(2)_s$ part of the parallel transporter ${\cal U}_{x,i}$.

In the Hilbert space of the theory local $SU(2)_s \otimes U(1)_Q$ 
transformations are implemented by unitary operators
\begin{equation}
V = \exp(i \sum_x \vec \eta(x) \cdot \vec S_x), \quad
W = \exp(i \sum_x \omega(x) Q_x),
\end{equation}
such that
\begin{eqnarray}
&&c_x' = V^\dagger c_x V = 
\exp(i \vec \eta(x) \cdot \frac{\vec \sigma}{2}) c_x = g(x) c_x, \quad 
g(x) \in SU(2)_s, \nonumber \\
&&^Qc_x = W^\dagger c_x W = \exp(i \omega(x)) c_x, \quad 
\exp(i \omega(x)) \in U(1)_Q.
\end{eqnarray}
Together with eqs.(\ref{SU2trafo}) and (\ref{U1trafo}) this implies that under 
the local transformations the Hamiltonian transforms as
\begin{equation}
H[{\cal U}'] = V H[{\cal U}] V^\dagger, \quad
H[^Q{\cal U}] = W H[{\cal U}] W^\dagger. 
\end{equation}
Similarly, one obtains
\begin{equation}
D_t' = V D_t V^\dagger, \quad ^QD_t = W D_t W^\dagger,
\end{equation}
such that the Schr\"odinger equation indeed transforms covariantly when one
uses
\begin{equation}
\Psi' = V \Psi, \quad ^Q\Psi = W \Psi.
\end{equation}

\subsection{External Electromagnetic Fields in the Effective Theory for 
Magnons and Charge Carriers}

The couplings of magnons to external electromagnetic fields have been 
investigated in detail in \cite{Bae04}. Again, the Fr\"ohlich-Studer symmetry 
is crucial and one obtains
\begin{equation}
\label{actionext}
S[\e,W_\mu] = \int d^2x \ dt \ \frac{\rho_s}{2} 
\left(D_i \e \cdot D_i \e + \frac{1}{c^2} D_t \e \cdot D_t \e\right),
\end{equation}
with the covariant derivative 
\begin{equation}
D_\mu \e(x) = \p_\mu \e(x) + \e(x) \times \vec W_\mu(x).
\end{equation}
Since magnons are electrically neutral, one may expect that they do not couple 
directly to the electromagnetic vector potential $A_\mu(x)$. Still, as 
discussed in \cite{Bae04} the issue is potentially non-trivial because there is
a Goldstone-Wilczek current
\begin{equation}
\label{jGW}
j_\mu^{GW}(x) = 
\frac{1}{8 \pi} \varepsilon_{\mu\nu\rho} \ \e(x) \cdot 
[D_\nu \e(x) \times D_\rho \e(x) + \vec W_{\nu\rho}(x)],
\end{equation}
with the non-Abelian field strength given by
\begin{equation}
\vec W_{\mu\nu}(x) = \p_\mu \vec W_\nu(x) - \p_\nu \vec W_\mu(x) -
\vec W_\mu(x) \times \vec W_\nu(x).
\end{equation}
The Goldstone-Wilczek current is an $SU(2)_s$ gauge-invariant extension of the
baby-Skyrmion current of eq.(\ref{baby}) and is also topologically conserved,
i.e.\ $\p_\mu j_\mu^{GW} = 0$. Hence, one may be tempted to add a 
Goldstone-Wilczek term $j_\mu^{GW}(x) A_\mu(x)$ to the Lagrangian. However, 
just like the Hopf term, the Goldstone-Wilczek term breaks $R$, $T$, and $T'$ 
and is thus forbidden in the present case.

Using the $P(x)$ notation, in the presence of external electromagnetic fields 
the action of eq.(\ref{actionext}) is given by
\begin{equation}
\label{actionextP}
S[P,W_\mu] = \int d^2x \ dt \ \rho_s \left(\mbox{Tr}[D_i P D_i P] +
\frac{1}{c^2} \mbox{Tr}[D_t P D_t P]\right),
\end{equation}
where the $SU(2)_s$ covariant derivative is denoted by
\begin{equation}
D_\mu P(x) = \p_\mu P(x) + [W_\mu(x),P(x)].
\end{equation}
As a consequence of the Fr\"ohlich-Studer symmetry, the action of 
eq.(\ref{actionextP}) is invariant even under local $SU(2)_s$ transformations
\begin{equation}
P(x)' = g(x) P(x) g(x)^\dagger, \quad 
W_\mu(x)' = g(x) (W_\mu(x) + \p_\mu) g(x)^\dagger.
\end{equation}

Let us now discuss how external electromagnetic fields enter the fermionic
part of the effective action. As a rule, ordinary derivatives must be replaced 
by covariant ones. This is the case also in the construction of the composite
vector field which now takes the form
\begin{equation}
\label{vSU2local}
v_\mu(x) = u(x) D_\mu u(x)^\dagger = u(x) [\p_\mu + W_\mu(x)] u(x)^\dagger.
\end{equation}
Under the local $SU(2)_s$ symmetry the field $u(x)$ transforms as
\begin{equation}
u(x)' = h(x) u(x) g(x)^\dagger,
\end{equation}
such that
\begin{eqnarray}
v_\mu(x)'&=&h(x) u(x) g(x)^\dagger [\p_\mu + g(x) (W_\mu(x) + \p_\mu) 
g(x)^\dagger] g(x) u(x)^\dagger h(x)^\dagger \nonumber \\
&=&h(x) u(x) [\p_\mu + W_\mu(x)] u(x)^\dagger h(x)^\dagger \nonumber \\
&=&h(x)(v_\mu(x) + \p_\mu) h(x)^\dagger.
\end{eqnarray}
This is exactly the same transformation behavior as for the global $SU(2)_s$ 
transformation of eq.(\ref{trafov}). In particular, this implies that the
$U(1)_s$ covariant derivative $D_\mu = \p_\mu + i v_\mu^3(x) \sigma_3$ need not
be modified when $SU(2)_s$ is turned into a local symmetry. Of course, 
according to eq.(\ref{vSU2local}), $v_\mu(x)$ now contains the electromagnetic 
fields $\vec E(x)$ and $\vec B(x)$ through the non-Abelian ``gauge'' field 
$W_\mu(x)$. Due to the local $U(1)_Q$ symmetry, the covariant derivatives still
need to be extended to
\begin{eqnarray}
&&D_\mu \Psi^{A,B} = \p_\mu \Psi^{A,B} + i v_\mu^3(x) \sigma_3 \Psi^{A,B} + 
\Psi^{A,B} i e A_\mu(x) \sigma_3, \nonumber \\
&&D_\mu \Psi^{A,B\dagger} = \p_\mu \Psi^{A,B\dagger} -
\Psi^{A,B\dagger} i v_\mu^3(x) \sigma_3 - 
i e A_\mu(x) \sigma_3 \Psi^{A,B\dagger}.
\end{eqnarray}
It should also be noted that the low-energy effective theory is not necessarily
just minimally coupled. In particular, the field strengths 
$F_{\mu\nu}(x) = \p_\mu A_\nu(x) - \p_\nu A_\mu(x)$ and $W_{\mu\nu}(x)$ may
also directly enter the low-energy effective theory.

\section{Conclusions}

We have constructed a systematic low-energy effective field theory describing
the interactions of magnons with charge carriers doped into an 
antiferromagnet. A key ingredient for constructing the effective theory are
symmetry considerations. The effective theory makes model-independent 
predictions for magnon-magnon, magnon-hole, and magnon-electron scattering. It
also determines the long-range magnon-mediated forces between electrons or 
holes. Although these would be highly non-trivial non-perturbative issues from 
the point of view of Hubbard-type models, in the framework of the effective
theory they can be understood quantitatively by perturbative analytic 
calculations. More ambitious non-perturbative questions might also be within 
reach of the effective theory. Such questions include the quantitative 
understanding of the Mott insulator state, the reduction of the staggered 
magnetization upon doping, the formation of a spiral phase, or the systematic 
investigation of dynamical mechanisms for the preformation of electron or hole 
pairs in the antiferromagnetic phase. In particular, magnon exchange --- the 
analog of pion exchange in nuclear physics --- suggests itself as a relevant 
mechanism.

Before one can do loop-calculations in the effective theory, one must 
establish a consistent power-counting scheme. This has originally been done
for pion chiral perturbation theory \cite{Gas85}, and carries over to magnon
chiral perturbation theory in a straightforward manner. When charge carriers
are included, the issue must be reconsidered. The same was true for
baryon chiral perturbation theory of pions and nucleons. In the baryon number
$B=1$ sector a consistent power-counting scheme enabling a systematic 
loop-expansion of the effective theory was established by Becher and Leutwyler
\cite{Bec99}. It is to be expected that this scheme can be extended to the
low-energy theory of magnons and charge carriers developed here. The
systematic power-counting in sectors with $B \geq 2$ still is a controversial 
issue in baryon chiral perturbation theory. The Weinberg power-counting scheme 
\cite{Wei90} seems to work in most (but not necessarily in all) cases. Its
relation to the alternative Kaplan-Savage-Wise scheme \cite{Kap98} should be 
clarified further \cite{Bea02,Nog05}. In light of the experience with effective
theories for the strong interactions, one should hence expect the issue of 
power-counting to be non-trivial in sectors with two or more charge carriers. 

Even when the extra $SU(2)_Q$ symmetry is imposed, in the fermion sector the 
effective theory has a large number of low-energy parameters. There are two 
rest mass parameters $M_1$ and $M_2$ as well as two kinetic mass parameters 
$M_1'$ and $M_2'$ for the fermions, four coupling constants $\Lambda_1$, 
$\Lambda_2$, $K_1$, and $K_2$ for fermion-one-magnon vertices, two coupling
constants $N_1$ and $N_2$ for fermion-two-magnon vertices, five 4-fermion
coupling constants $G_1,G_2,...,G_5$, two 6-fermion couplings $H_1$ and $H_2$, 
and finally one 8-fermion coupling $I$. If only the $U(1)_Q$ symmetry is 
imposed there are even more parameters. The large number of a priori 
undetermined low-energy parameters is the price one has to pay for the 
universality and model-independence of the effective theory. Only in this way 
the low-energy physics of any arbitrary cuprate antiferromagnet can be captured
by the effective theory. Of course, due to the rather large number of 
parameters, the predictive power of the effective theory is somewhat limited. 
Still, only a few parameters enter in some relevant physical quantities. For 
example, the one-magnon exchange potential between charge carriers depends only
on certain combinations of the fermion-magnon couplings $\Lambda_1$, 
$\Lambda_2$, $K_1$, and $K_2$. Also, for example, the details of the 
short-range 4-, 6-, and 8-fermion couplings are not expected to be essential 
for identifying potential mechanisms for preforming electron or hole pairs in 
the antiferromagnetic phase. It is interesting to note that the low-energy 
effective theory of the $t$-$J$ model, in which electrons are excluded beyond 
half-filling and holes are the only charge carriers, has a much smaller number 
of low-energy parameters. In that case, there are only one rest mass parameter 
$M$, one kinetic mass parameter $M'$, two coupling constants $\Lambda$ and $K$ 
for hole-one-magnon vertices, one coupling constant $N$ for a hole-two-magnon 
vertex, and one 4-fermion coupling constant $G$. It would be interesting to 
perform numerical simulations of the Hubbard or $t$-$J$ model in order to 
determine the values of the corresponding low-energy parameters by comparison 
with calculations in the effective theory. For example, in the $t$-$J$ model 
one can determine the parameters $M$, $M'$, $\Lambda$, $K$, and $N$ from 
simulations in the one-hole sector, while the determination of $G$ requires 
computations in the two-hole sector of the Hilbert space.

It should be pointed out that, as it stands, the effective theory is applicable
only at small doping, i.e.\ for small $\mu$. This is sufficient for 
understanding the long-range forces between electrons or holes in the 
antiferromagnetic phase. It should also allow a quantitative investigation of
the reduction of the staggered magnetization upon doping. However, in order to 
enter the high-temperature superconducting phase itself, if this is at all
possible within the effective field theory presented here, larger values of 
$\mu$ will be necessary. Once $\mu$ becomes large, it sets a new scale which 
must be taken into account in the power-counting. However, most important, the 
symmetry considerations of the present paper still apply in that case as well.

Some of the most interesting questions one can address in the framework of the
effective theory may require non-perturbative calculations. While in some cases
such calculations can be performed in the continuum, in others they may require
a non-perturbative regularization of the effective theory. In \cite{Cha03} the
effective theory of pions and nucleons was regularized on a space-time lattice 
in order to address non-perturbative questions concerning the strong 
interactions. It may also be useful to formulate the effective theory of 
magnons and charge carriers on the lattice. For example, it would be 
interesting to investigate if the effective theory is more easily solvable by 
numerical simulation than the standard Hubbard-type models.

We like to emphasize again that effective field theory also allows us to 
include phonons in addition to magnons. This may shed light on more complicated
potential mechanisms for Cooper pair preformation which involve both magnon and
phonon exchange. It is interesting to construct such an effective theory. In 
particular, the Galilean (or even Poincar\'e) symmetry is then non-linearly 
realized.

To summarize, low-energy effective field theory is a powerful tool that 
has several advantages compared to the direct use of microscopic models. First,
it is model-independent and provides universal predictions. Material-specific 
details of the underlying microscopic system enter the effective theory only 
through low-energy parameters whose values can be determined by comparison with
experiments or with numerical simulations. Second, and most important, the 
electrons or holes of the effective theory are quasi-particles whose long-range
forces are weak and calculable in perturbation theory. This is a significant 
advantage compared to calculations in microscopic models of strongly correlated
electrons which are necessarily non-perturbative. While it is practically 
impossible to reliably determine the long-range forces between charge carriers 
from Hubbard-type models, in the effective theory the calculation of the 
one-magnon exchange forces is straightforward and presently in progress. It is 
very interesting to ask if these forces will provide a potential mechanism for 
the preformation of electron or hole pairs. In any case, we propose the
systematic low-energy effective field theory approach as a better compromise 
between calculability and predictive power than the one offered by Hubbard-type
models. Effective field theory sheds new light on the dynamics of charge 
carriers in antiferromagnets, and there is hope that it may even be applicable 
to the high-temperature superconductors themselves.

\section*{Acknowledgements}

We have benefitted from discussions with M.\ Bissegger, S.\ Chandrasekharan,
G.\ Colangelo, J.\ Gasser, P.\ Hasenfratz, H.\ Leutwyler, P.\ Minkowski, and
F.\ Niedermayer. This work is supported by funds provided by the 
Schweizerischer Nationalfonds.

\begin{appendix}

\section{Electron-Hole Representation of the Hubbard Model Operators}

For $U \gg |t|$ the Hubbard model at half-filling reduces to the 
antiferromagnetic quantum Heisenberg model. In contrast to the Heisenberg 
ferromagnet, the ground state of the antiferromagnet is not known
analytically. In particular, the naive N\'eel state 
\begin{equation}
|N\rangle = \prod_{x \in A} c^\dagger_{x \downarrow} 
\prod_{x\in B} c^\dagger_{x \uparrow} |0\rangle,
\end{equation} 
with all spins down on the even sublattice $A$ and all spins up on the odd
sublattice $B$ is not an eigenstate of the Hubbard Hamiltonian. Still, we use 
this state in order to define electron and hole operators. For even sites we
then find
\begin{equation}
c_{x \uparrow} |N\rangle = 0, \quad c^\dagger_{x \downarrow} |N\rangle = 0, 
\quad x \in A.
\end{equation}
Correspondingly, $c^\dagger_{x \uparrow}$ creates an electron, while 
$c_{x \downarrow}$ creates a hole. Hence, just like a relativistic Dirac 
spinor, the $SU(2)_s$ spinor
\begin{equation}
c_x = \left(\begin{array}{c} c_{x \uparrow} \\ 
c_{x \downarrow} \end{array}\right) = \left(\begin{array}{c} a_{x \uparrow} \\ 
b_{x \uparrow}^\dagger \end{array}\right), \quad  x \in A,
\end{equation}
consists of a particle annihilation operator $a_{x \uparrow}$ in the upper 
component and a hole creation operator $b_{x \uparrow}^\dagger$ in the lower 
component. Note that the annihilation of an electron with spin down via 
$c_{x \downarrow}$ corresponds to the creation of a hole with spin up via
$b_{x \uparrow}^\dagger$. Similarly, on the odd sites one has
\begin{equation}
c_{x \downarrow} |N\rangle = 0, \quad c^\dagger_{x \uparrow} |N\rangle = 0, 
\quad x \in B.
\end{equation}
In this case, $c^\dagger_{x \downarrow}$ creates a particle, while 
$c_{x \uparrow}$ creates a hole and we write
\begin{equation}
c_x = \left(\begin{array}{c} c_{x \uparrow} \\ 
c_{x \downarrow} \end{array}\right) = 
\left(\begin{array}{c} b_{x \downarrow}^\dagger \\ 
a_{x \downarrow} \end{array}\right), \quad  x \in B.
\end{equation}

\section{Removal of Non-Canonical Terms by a Field Redefinition}

The most general $SU(2)_Q$-breaking but $U(1)_Q$-symmetric terms containing one
covariant time-derivative are given by
\begin{eqnarray}
&&\frac{a}{2} \mbox{Tr}[\Psi^{A\dagger} D_t \Psi^A +
\Psi^{B\dagger} D_t \Psi^B] +
\frac{b}{2} \mbox{Tr}[\Psi^{A\dagger} \sigma_3 D_t \Psi^A \sigma_3 - 
\Psi^{B\dagger} \sigma_3 D_t \Psi^B \sigma_3] \nonumber \\
&&+ \frac{c}{2} \mbox{Tr}[\Psi^{A\dagger} D_t \Psi^B \sigma_3 + 
\Psi^{B\dagger} D_t \Psi^A \sigma_3] \nonumber \\
&&\hspace{2cm}= (\psi_+^{A\dagger}, \psi_+^{B\dagger}) \left(\begin{array}{cc}
a + b & c \\ c & a - b \end{array} \right) \left(\begin{array}{c}
D_t \psi_+^A \\ D_t \psi_+^B \end{array} \right) \nonumber \\
&&\hspace{2cm} \phantom{=}
+ (\psi_-^{A\dagger}, \psi_-^{B\dagger}) \left(\begin{array}{cc}
a - b & c \\ c & a + b \end{array} \right) \left(\begin{array}{c}
D_t \psi_-^A \\ D_t \psi_-^B \end{array} \right) \nonumber \\
&&\hspace{2cm}= (\widetilde\psi^{A\dagger}_+,\widetilde\psi^{B\dagger}_+)
\left(\begin{array}{c} D_t \widetilde \psi_+^A \\ 
D_t \widetilde\psi_+^B \end{array} \right) +
(\widetilde\psi^{A\dagger}_-,\widetilde\psi^{B\dagger}_-)
\left(\begin{array}{c} D_t \widetilde\psi_-^A \\ 
D_t \widetilde\psi_-^B \end{array} \right).
\end{eqnarray}
Here $\widetilde\psi^{A,B}_\pm(x)$ results from a field redefinition that
diagonalizes the matrices in the previous expression. Only the term 
proportional to $a$ contains the standard form 
$\psi^{A\dagger}_+ \p_t \psi^A_+ + \psi^{A\dagger}_- \p_t \psi^A_- + 
\psi^{B\dagger}_+ \p_t \psi^B_+ + \psi^{B\dagger}_- \p_t \psi^B_-$
which implies canonical anticommutation relations between fermionic creation 
and annihilation operators in the Hamiltonian formulation. The non-canonical 
terms (proportional to $b$ and $c$) can be removed by an appropriate field 
redefinition
\begin{eqnarray}
&&\left(\begin{array}{c}\widetilde\psi^A_\pm(x) \\ \widetilde\psi^B_\pm(x) 
\end{array}\right) = \left(\begin{array}{cc} \sqrt{\lambda_\pm} & 0 \\ 
0 & \sqrt{\lambda_\mp} \end{array} \right) U_\pm 
\left(\begin{array}{c} \psi^A_\pm(x) \\ \psi^B_\pm(x) \end{array} \right), 
\nonumber \\
&&\lambda_\pm = a \pm \sqrt{b^2 + c^2}, \quad
U_\pm = \left(\begin{array}{cc} X & \pm Y \\ \mp Y & X \end{array} \right).
\end{eqnarray}
Here $U_\pm$ are unitary matrices with $X, Y \in \R$ which obey
\begin{equation}
U_\pm \left(\begin{array}{cc} a \pm b & c \\ c & a \mp b \end{array} 
\right)
U_\pm^\dagger = \left(\begin{array}{cc} \lambda_\pm & 0 \\
0 & \lambda_\mp \end{array} \right).
\end{equation}

It is straightforward to show that the redefined fields 
$\widetilde\psi^{A,B}_\pm(x)$ have the same symmetry properties of 
eqs.(\ref{psitrafo1}) and (\ref{psitrafo2}) as the original fields 
$\psi^{A,B}_\pm(x)$. Under the $SU(2)_s$ symmetry the original fields transform
as
\begin{equation}
\psi^{A,B}_\pm(x)' = \exp(\pm i \alpha(x)) \psi^{A,B}_\pm(x),
\end{equation}
and after the field redefinition again
\begin{eqnarray}
\!\!\!\!\!\!\widetilde\psi^A_\pm(x)'&=&\sqrt{\lambda_\pm}
[X \psi^A_\pm(x)' \pm Y \psi^B_\pm(x)'] \nonumber \\
&=&\exp(\pm i \alpha(x)) \sqrt{\lambda_\pm}
[X \psi^A_\pm(x) \pm Y \psi^B_\pm(x)] =
\exp(\pm i \alpha(x)) \widetilde\psi^A_\pm(x), \nonumber \\
\!\!\!\!\!\!\widetilde\psi^B_\pm(x)'&=&\sqrt{\lambda_\mp}
[\mp Y \psi^A_\pm(x)' + X \psi^B_\pm(x)'] \nonumber \\
&=&\exp(\pm i \alpha(x)) \sqrt{\lambda_\mp}
[\mp Y \psi^A_\pm(x) + X \psi^B_\pm(x)] =
\exp(\pm i \alpha(x)) \widetilde\psi^B_\pm(x).
\end{eqnarray}
Similarly, under the $U(1)_Q$ symmetry the original fields transform as
\begin{equation}
^Q\psi^{A,B}_\pm(x) = \exp(i \omega) \psi^{A,B}_\pm(x),
\end{equation}
and again
\begin{eqnarray}
^Q\widetilde\psi^A_\pm(x)&=&\sqrt{\lambda_\pm}
[X \ ^Q\psi^A_\pm(x) \pm Y \ ^Q\psi^B_\pm(x)] \nonumber \\
&=&\exp(i \omega) \sqrt{\lambda_\pm}
[X \psi^A_\pm(x) \pm Y \psi^B_\pm(x)] =
\exp(i \omega) \widetilde\psi^A_\pm(x), \nonumber \\
^Q\widetilde\psi^B_\pm(x)&=&\sqrt{\lambda_\mp}
[\mp Y \ ^Q\psi^A_\pm(x) + X \ ^Q\psi^B_\pm(x)] \nonumber \\
&=&\exp(i \omega) \sqrt{\lambda_\mp}
[\mp Y \psi^A_\pm(x) + X \psi^B_\pm(x)] =
\exp(i \omega) \widetilde\psi^B_\pm(x).
\end{eqnarray}
Under the modified displacement symmetry $D'$ one has
\begin{equation}
^{D'}\psi^{A,B}_\pm(x) = \pm \psi^{B,A}_\mp(x),
\end{equation}
and after the field redefinition one again obtains 
\begin{eqnarray}
^{D'}\widetilde\psi^A_\pm(x)&=&\sqrt{\lambda_\pm}
[X \ ^{D'}\psi^A_\pm(x) \pm Y \ ^{D'}\psi^B_\pm(x)] \nonumber \\
&=&\pm \sqrt{\lambda_\pm}
[X \psi^B_\mp(x) \pm Y \psi^A_\mp(x)] = \pm \widetilde\psi^B_\mp(x), 
\nonumber \\
^{D'}\widetilde\psi^B_\pm(x)&=&\sqrt{\lambda_\mp}
[\mp Y \ ^{D'}\psi^A_\pm(x) + X \ ^{D'}\psi^B_\pm(x)] \nonumber \\
&=&\pm \sqrt{\lambda_\mp}
[\mp Y \psi^B_\mp(x) + X \psi^A_\mp(x)] = \pm \widetilde\psi^A_\mp(x).
\end{eqnarray}
Since the displacement symmetry $D$ is a combination of $D'$ and $SU(2)_s$ it
also maintains its original form. The same is true for the discrete symmetries
$O$ and $R$. Finally, under the modified time-reversal $T'$ the original fields
transform as
\begin{equation}
^{T'}\psi^{A,B}_\pm(x) = - \psi^{A,B\dagger}_\pm(Tx),
\end{equation}
such that
\begin{eqnarray}
^{T'}\widetilde\psi^A_\pm(x)&=&\sqrt{\lambda_\pm}
[X \ ^{T'}\psi^A_\pm(x) \pm Y \ ^{T'}\psi^B_\pm(x)] \nonumber \\
&=&- \sqrt{\lambda_\pm}
[X \psi^{A\dagger}_\pm(Tx) \pm Y \psi^{B\dagger}_\pm(Tx)] = 
- \widetilde\psi^{A\dagger}_\pm(Tx), \nonumber \\
^{T'}\widetilde\psi^B_\pm(x)&=&\sqrt{\lambda_\mp}
[\mp Y \ ^{T'}\psi^A_\pm(x) + X \ ^{T'}\psi^B_\pm(x)] \nonumber \\
&=&- \sqrt{\lambda_\mp}
[\mp Y \psi^{A\dagger}_\pm(Tx) + X \psi^{B\dagger}_\pm(Tx)] = 
- \widetilde\psi^{B\dagger}_\pm(Tx).
\end{eqnarray}
As a combination of $T'$ and $SU(2)_s$ the time-reversal symmetry $T$ also
maintains its original form after the field redefinition. The only symmetry 
that does not maintain its original form is $SU(2)_Q$. This is no 
problem since the non-canonical terms can arise only when the $SU(2)_Q$ 
symmetry is explicitly broken down to $U(1)_Q$ and is hence no longer a 
symmetry of the theory.

Since the redefined fields transform exactly like the original ones, the terms 
in the effective Lagrangian take exactly the same form as before. Hence, it is 
indeed justified not to include the non-canonical terms in the effective 
Lagrangian.

\end{appendix}

\end{document}